\documentclass[aps,prb,twocolumn,superscriptaddress,showpacs,10pt]{revtex4-1} 

\usepackage{graphicx}
\usepackage{color}
\usepackage{latexsym}
\usepackage{amsmath}
\usepackage{amsfonts}
\usepackage{amssymb}
\usepackage{subfigure}
\usepackage{bm}
\usepackage{hyperref}
\hypersetup{colorlinks=true}
\usepackage[all]{hypcap} 

\begin{document}

\title{
Superradiant-like dynamics by electron shuttling on a nuclear-spin island
}

\author{Yi-Nan Fang} 
\affiliation{Beijing Computational Science Research Center, Beijing 100193, China}
\affiliation{Department of Physics, McGill University, 3600 Rue University, Montreal, Quebec, Canada H3A 2T8}

\author{Ying-Dan Wang}
\affiliation{CAS Key Laboratory of Theoretical Physics, Institute of Theoretical Physics, Chinese Academy of Sciences, Beijing 100190, China}
\affiliation{School of Physical Sciences \& CAS Center for Excellence in Topological Quantum Computation, University of Chinese Academy of Sciences, Beijing 100049, China}
\affiliation{Synergetic Innovation Center for Quantum Effects and Applications, Hunan Normal University, Changsha 410081, China}

\author{Rosario Fazio} 
\affiliation{ICTP, Strada Costiera 11, I-34151 Trieste, Italy}
\affiliation{Dipartimento di Fisica, Universita' di Napoli ``Federico II'', Monte S. Angelo, I-80126 Napoli, Italy}

\author{Stefano Chesi}
\email{stefano.chesi@csrc.ac.cn}
\affiliation{Beijing Computational Science Research Center, Beijing 100193, China}
\affiliation{Department of Physics, Beijing Normal University, Beijing 100875, China}

\begin{abstract}
We investigate superradiant-like dynamics of the nuclear-spin bath in a single-electron quantum dot, by considering electrons cyclically shuttling on/off an isotopically enriched `nuclear-spin island'. Assuming a uniform hyperfine interaction, we discuss in detail the nuclear spin evolution under shuttling and its relation to superradiance. We derive the minimum shuttling time which allows to escape the adiabatic spin evolution. Furthermore, we discuss slow/fast shuttling under the inhomogeneous field of a nearby micromagnet. Finally, by comparing our scheme to a model with stationary quantum dot, we stress the important role played by non-adiabatic shuttling in lifting the Coulomb blockade and thus establishing the superradiant-like behavior. 
\end{abstract}

\date{\today}

\maketitle

Coherent control of spins in solid-state systems is a subject of intense research, both from the point view of fundamental physics as well as future applications. In quantum dots, significant efforts have been directed towards understanding the coupling of electronic spins to the nuclear-spin bath of the semiconductor host (see, e.g., Refs.~\onlinecite{2009_Phys_Stat_Sol_b_Coish,2017_Rep_Phys_Yang}, and references therein). Remarkably, a stochastic classical description of the nuclear (Overhauser) field has proved very useful in modeling decoherence at short time scales,\cite{2007_PRL_Koppens,2016_PRL_Chesi}  developing efficient dynamical decoupling techniques,\cite{2011_Nat_Phys_Bluhm,2016_Nat_Nano_Malinowski} and suppressing nuclear noise through feedback-loops or postselection.\cite{2010_PRL_Bluhm,2014_Nat_Commun_Shulman,2016_PRL_Delbecq,2017_PRL_Malinowski} The observation and/or prediction of quantum phenomena which rely on the \emph{coherent} nature of the electron-nuclear interaction is also an interesting objective. An example of this sort is the precise control of the electron-nuclear system of impurity centers, leading to long-lived storage of quantum information\cite{2013_Nature_Pla} and the realization of small quantum registers.\cite{2012_Nature_van_der_Sar} 

With quantum dots, which typically host a dense distribution of up to $N \sim 10^5 -10^6$ nuclear spins, addressing individual nuclear spins is much more challenging. A line of theoretical research has been guided by the analogy of the uniform-coupling limit of the electron-nuclear Hamiltonian to the Dicke model of optical superradiance,\cite{1954_PR_Dicke,1971_PRA_Degiorgio,1982_Phys_Rep_Gross,2010_PRL_Kessler,2019_PRB_He} and focused on the generation of large-scale nuclear-spin coherence through collectively enhanced electron-nuclear spin flips.\cite{2004_JPSJ_Eto,2010_PRL_Kessler,2012_PRB_Schuetz,2015_PRB_Stefano} An attractive feature of these proposals is that the collective enhancement would be proportional to $N \gg 1$. Here we investigate the possibility of realizing the superradiant-like enhancement in a movable quantum dot configuration, where the electron is shuttled between two external reservoirs.\cite{1998_PRL_Gorelik,1998_Physica_B_Gorelik,1998_Physica_B_Isacsson,1998_PU_Gorelik,2001_Nature_Gorelik} As we will see, such a shuttling device offers special advantages in the realization of superradiant-like evolution. Further motivations come from recent experimental progress on shuttling electrons across extended quantum dot arrays.\cite{2017_npj_Quant_Info_Fujita,2018_Petta}  More generally, electron shuttles can also be realized in nano-electro-mechanical systems with vibrating organic molecules,\cite{2000_Nature_Park} metallic grains,\cite{1998_PRL_Gorelik} or silicon nanopillars,\cite{2004_APL_Scheible} and are characterized by rich transport regimes due to the interplay of charge and mechanical degree of freedoms.\cite{2004_PRL_Novotny,2005_PRL_Pistolesi,2005_NJP_Donarini} They also attract interest in the study of noise and full counting statistics.\cite{2004_PRB_Pistolesi,2004_PRB_Romito}

\begin{figure}
\begin{centering}
\includegraphics[width=0.48\textwidth]{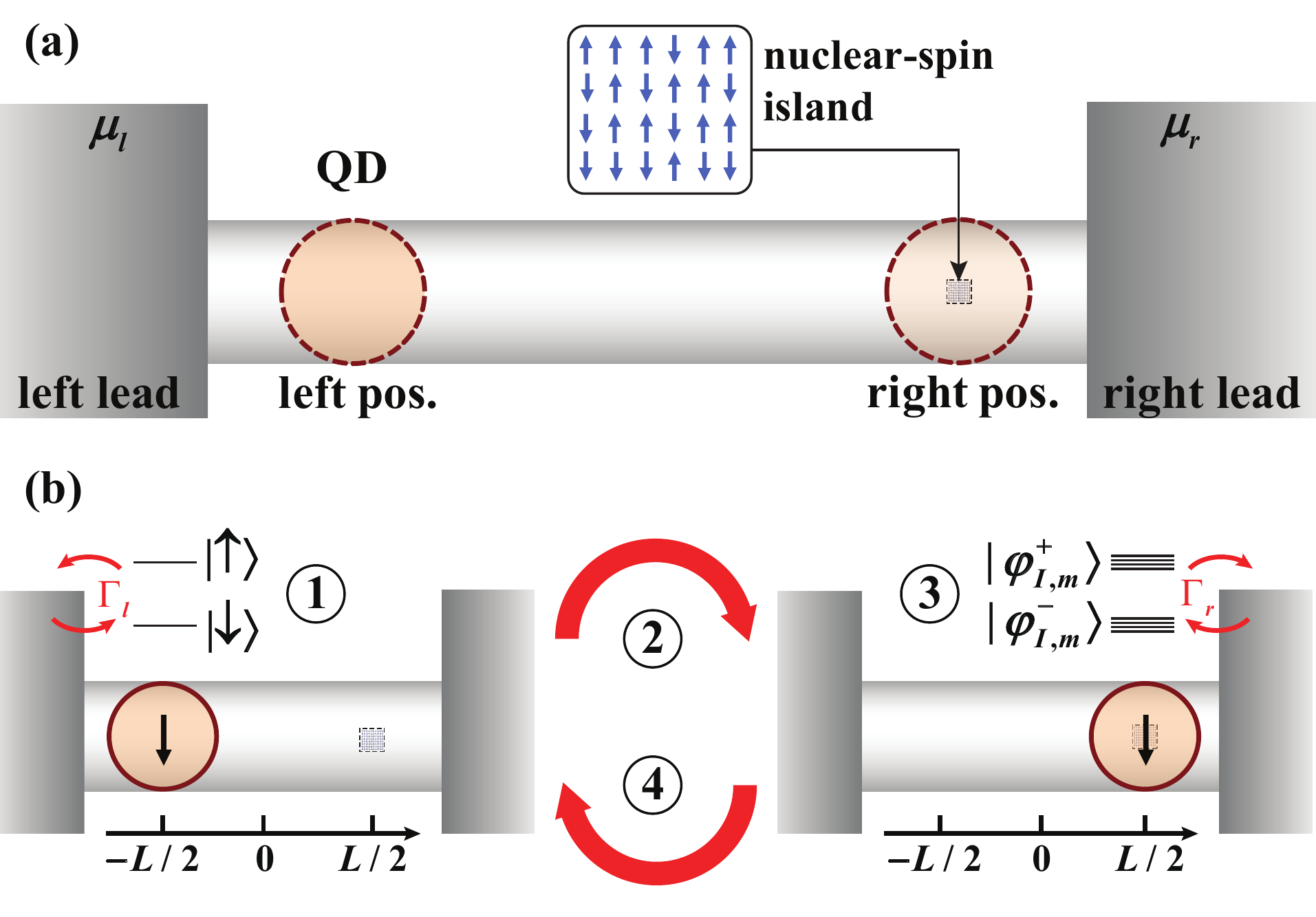}
\end{centering}
\caption{(a) Schematics of the electron shuttle. An excess electron resides on the orbital ground state of the instantaneous trapping potential, forming a moving quantum dot (QD). The trapping potential can be controlled between a nuclear-spin free region (left position, at $x=-L/2$) and a nuclear-spin rich region (right position, at $x=L/2$).\cite{2015_PRB_Stefano}  The size of the QD envelope function is significantly larger than the nuclear region, allowing a ``box-model'' description for their coherent coupling.\cite{2003_PRB_Khaetskii,2007_JAP_Coish,2006_PRB_Zhang} (b) Shuttling cycle. In step 1 (step 3) the dot is tunnel-coupled to the left (right) external lead, with energy levels as schematically illustrated.  Steps 2 and 4 are fast shuttling processes between $x=\pm L/2$.} \label{schematics}
\end{figure}

In our setup, schematically illustrated in in Fig.~\ref{schematics}, an electron is trapped in a quantum dot whose center position $x(t)$ can be controlled via external gates (e.g., along a nanowire). A shuttling motion is imposed between left and right operating points, which are in contact with external leads. Furthermore, a nuclear-spin rich region is embedded at the right position and the periodic interaction with nuclear spins is able to induce an interesting interplay between the charge and spin degrees of freedom. While at the left position ($x=-L/2$, poor in nuclear spins) the hyperfine interaction is effectively turned off, on top of the spatially localized `nuclear-spin island' ($x=L/2$) the system approaches the ideal limit of nuclear spins with nearly equal hyperfine strength.\cite{2015_PRB_Stefano} This condition leads to a simple integrable Hamiltonian which is in direct analogy to the (infinite range) Dicke model. Shuttling between the two operating points allows to separate spatially the entangled electron-nuclear dynamics from the electron-spin initialization along the external magnetic field, thus implementing the superradiant-like dynamics in a rather direct manner.

Our article is organized as follows: In Sec.~\ref{sec:models} we present the electron shuttle model. In Sec.~\ref{sec:evolution} the combined electron-nuclear spin dynamics is analyzed under the assumption of fast shuttling. In Sec.~\ref{stochastic_evolution} we present an alternative analysis in terms of Monte Carlo wave-function simulations, which allows to discuss the signatures of superradiant-like dynamics in charge sensing and current fluctuation. In Sec.~\ref{sec:initialization} we derive the non-adiabaticity condition for the spin evolution (depending on shuttling speed). A strictly related discussion of shuttling in the slanting field of a micromagnet is also provided. In Sec.~\ref{sec:comparison}, we discuss the crucial role played by the non-adiabatic shuttling in weak-tunneling setups, as it allows to lift the Coulomb blockade regime and induce the desired superradiant-like evolution. Further technical details can be found in Appendices~\ref{section_1a} and \ref{sec_rates}.

\section{The model}\label{sec:models}

The shuttling setup studied in this paper is schematically illustrated in Fig.~\ref{schematics}. We model it as a moving quantum dot, whose time-dependent position $x(t)$ (i.e., the minimum of the confining potential) can be controlled externally. To specify the Hamiltonian, it is convenient to consider first a given value of $x$, which fixes the couplings at their instantaneous value. We will  describe later the shuttling cycle and the associated electron and nuclear-spin dynamics.

\subsection{Hyperfine interaction and tunnel couplings}

We suppose that the shuttling is sufficiently slow, such that during the whole process the electrons occupy the instantaneous orbital ground state of the quantum dot. Furthermore, due to a large Coulomb repulsion, we neglect doubly-occupied states. At a given value of $x$, the singly-occupied states are $d_\sigma^\dag |0\rangle$, where $|0\rangle$ is the state with no electrons in the dot, $d_\sigma^\dag$ is a fermionic creation operator, and $\sigma=\uparrow,\downarrow$ is the spin index. The full Hamiltonian reads:
\begin{equation}\label{H}
H=H_{0}+H_{T}+H_{b},
\end{equation}
where the isolated dot is described by ($\hbar =1$): 
\begin{equation}\label{H_qdot}
H_{0}=\sum_{\sigma}\epsilon_{\sigma}d_{\sigma}^{\dagger}d_{\sigma}+\frac{A}{N_d} {\bf S} \cdot \bf{I},
\end{equation}
where $\omega_{0}=\epsilon_{\uparrow}-\epsilon_{\downarrow}=g\mu_{B}B_{z}$ is the Zeeman splitting due to an external magnetic field in the $z$ direction and the second term is the hyperfine interaction, with $S^\alpha = \frac12\sum_{ss'} \sigma_{ss'}^\alpha d_{s}^{\dagger}d_{s'}$ ($\boldsymbol{\sigma}$ is the vector of Pauli matrices) the electron spin operators and $I^{\alpha}=\sum_{i=1}^{N}I_{i}^{\alpha}$ the collective spin operators of $N$ nuclear spins. 

In general, the coupling strength of the hyperfine interaction for a nuclear spin at position ${\bf r}_k$ has the form $A  v_0 |\psi({\bf r}_k)|^2$, where the energy scale $A$ depends on the nuclear isotope and the electronic states of the host crystal, $v_0$ is the atomic volume, and $\psi({\bf r})$ is the envelope function of the quantum dot.\cite{2008_Adv_Phys_Chirolli,2009_PSSB_Coish,2007_JAP_Coish} Here we have approximated $A  v_0 |\psi({\bf r}_k)|^2\simeq A/N_d$, which is justified under special circumstances. For example, Ref.~\onlinecite{2015_PRB_Stefano} proposed to realize approximately uniform hyperfine couplings through a `nuclear-spin island'. As discussed there, the concept might be implemented in a Si/Ge core-shell nanowire with a segment of its inner core being isotopically modulated to host a $^{29}$Si section of nanometer size.\cite{2002_Nature_Lauhon,2011_APL_Moutanabbir}  Alternatively, the right position $x=L/2$ could host one or few magnetic impurities.\cite{2015_PRB_Lai} We note that $N_d$ is of the order of the number of lattice sites having significant overlap with the quantum dot. Thus, for materials with spinless isotopes, $N$ can be significantly smaller than $N_d$.

Taking into account the nuclear spins, the empty quantum dot is simply described by $|0,m\rangle \equiv |0\rangle \otimes |I, m \rangle$, where $|I, m \rangle$ are the eigenstates of $I^2, I^z$ with eigenvalues $I(I+1)$ and $m$, respectively (we omit a permutational quantum number). In the basis  $|\sigma,m\rangle \equiv d_\sigma^\dag |0\rangle \otimes |I,m \rangle$, the eigenstates with one electron are given by:
\begin{align}\label{eigenstates}
&|\varphi_{I,m}^{-}\rangle=\alpha_{m-1}|\downarrow,m\rangle-\beta_{m-1}|\uparrow,m-1\rangle, \nonumber \\
&|\varphi_{I,m}^{+}\rangle=\alpha_m|\uparrow,m\rangle+\beta_m|\downarrow,m+1\rangle,
\end{align}
where $m \in [-I,I]$ and, conventionally, $|\uparrow ,-I-1 \rangle = |\downarrow, I+1 \rangle =0$. The amplitudes are $\alpha_{m}=\cos(\theta_{m}/2)$ and $\beta_{m}=\sin(\theta_{m}/2)$, with the mixing angle:
\begin{equation}\label{theta_m}
\theta_{m}=\arg\left[\frac{1}{2\eta}+m+\frac12+i\sqrt{I(I+1)-m(m+1)}\right].
\end{equation}
The parameter $\eta$ is the ratio of hyperfine coupling and Zeeman energy:
\begin{equation}
\eta=\frac{A/N_d}{2\omega_0},
\end{equation}
and will play an important role in the rest of the paper. For typical quantum dots, $\eta \ll  1 $ under a moderate magnetic field and we will also restrict ourselves to this condition. 
For example, using values appropriate to Si quantum dots\cite{Perry_arXiv} $A \simeq 2~\mu{\rm eV}$,  $\omega_0 = 10~{\mu \rm eV}$ (i.e., $B_z \simeq 0.1$~T), and $N_d = 10^5$, one obtains $\eta \simeq  10^{-6} $. Finally, the energies of $|\varphi_{I,m}^{\pm}\rangle$ are:
\begin{align}
\epsilon^{\pm}_{I,m}=\bar{\epsilon} \pm\frac{\omega_0}{2}\sqrt{1+\eta(4m \pm 2)+\eta^2(2I+1)^2}- \frac{\eta\omega_0}{2} ,
\end{align}
where we defined $\bar{\epsilon} =(\epsilon_\uparrow+\epsilon_\downarrow)/2 $. If the condition  $\eta I \ll 1$ is satisfied, the $\pm$ sets of levels form two energy bands separated by a large gap close to $\omega_0$. We choose the level alignment as in Fig.~\ref{schematics}(b), where $\epsilon_\uparrow \sim \epsilon^{+}_{I,m} > \mu_{l,r} > \epsilon_\downarrow \sim \epsilon^{-}_{I,m} $. 

The quantum dot is connected to two external leads through a standard tunnel Hamiltonain:
\begin{equation}\label{H_T}
H_{T}=\sum_{\alpha,k,\sigma}T_{\alpha k}d_{\sigma}^{\dagger}c_{\alpha k\sigma}+\mathrm{H.c.},
\end{equation}
with spin-independent tunnel amplitudes, $T_{\alpha k}$ and $\alpha=l,r$ labeling the left and right lead, respectively. $H_b$ is:
\begin{equation}
H_{b}=\sum_{\alpha,k,\sigma}\varepsilon_{\alpha k}c_{\alpha k\sigma}^{\dagger}c_{\alpha k\sigma},
\end{equation}
where we assume that the reservoirs are unpolarized, thus the single-particle energies $\varepsilon_{\alpha k}$ are spin-independent. As a consequence, the density of states $n_\alpha(\varepsilon)$ are spin-independent as well. The occupation numbers are given by $f_{\alpha}(\varepsilon)=\{\exp[\beta(\varepsilon-\mu_{\alpha})]+1\}^{-1}$ where we generally assume the low-temperature regime:
\begin{equation}\label{f_leads}
f_{\alpha}(\varepsilon) \simeq \theta(\mu_\alpha - \varepsilon).
\end{equation}
Although other choices are possible, the desired spin dynamics can be generated without an applied bias. Therefore we will assume $\mu_l = \mu_r$, as illustrated in Fig.~\ref{schematics}(b).

\subsection{Electron shuttle}\label{shuttling_setup}

While in some shuttling setups it is necessary to solve a separate dynamical equation for the moving center $x(t)$, depending on the evolution of the internal variables of the shuttle (e.g., its charge state\cite{1998_PRL_Gorelik}), here we assume that the motion is determined by external controls. In particular, we neglect the small back-action on the electron motion from its internal spin dynamics. The main consequence on the system Hamiltonian Eq.~(\ref{H}) of the $x(t)$ parametric dependence is to induce time-dependent tunnel and hyperfine couplings.

As represented in Fig.~\ref{schematics}, the right and left operating points are at $x_l=- L/2$ and $x_r = L/2$, respectively. When the electron shuttle moves close to $x_l$ ($x_r$) it interacts more strongly with the left (right) lead. We can write explicitly the $x$-dependence of the tunnel amplitudes in Eq.~(\ref{H_T}) as follows:
\begin{equation}
T_{\alpha k}(x) \simeq T_{\alpha}e^{-|x-x_\alpha|/\lambda_{\alpha}},
\end{equation}
where $\lambda_{l,r}$ are the tunneling lengths.\cite{1998_PRL_Gorelik,2003_J_Phys_Cond_Mat_Shekhter,2005_NJP_Donarini} Here we have also made the usual approximations that $T_{\alpha k}$ is independent of $k$. Further assuming $n_{\alpha}(\varepsilon)\simeq n_\alpha$, the tunneling rates at the left/right positions are 
\begin{equation}\label{Gamma_lr}
\Gamma_{\alpha}=2\pi n_{\alpha}|T_{\alpha}|^{2},
\end{equation} 
which we choose to be in the weak-tunneling regime, $\Gamma_\alpha \ll |\mu_\alpha - \epsilon_{I,m}^{\pm} | $. For simplicity, we will also consider $\lambda_{l,r}\ll L$, such that an electron at $x_l$ ($x_r$) can only interact with the left (right) reservoir. 

Similarly, the spatial dependence of the hyperfine interaction could be of the type:
\begin{equation}\label{Ax}
A(x)=A e^{-(L/2-x)^2/\Delta x^2},
\end{equation}
where we take into account a Gaussian envelope wavefunction (appropriate for a harmonic confinement centered in $x$). To have all the hyperfine couplings approximately equal, the spatial extent of the nuclear-spin rich region should be smaller than $\Delta x$. Furthermore, we will typically assume $\Delta x \ll L$ such that the hyperfine coupling is only significant when $x\simeq L/2$. The assumption of uniform coupling is more accurate when the center of the electron's wavefunction sits on top of the small nuclear-spin island.\cite{2015_PRB_Stefano} At this position, the hyperfine coupling is largest.

\section{Superradiant-like shuttling}\label{sec:evolution}

We now consider the electron-nuclear spin dynamics under a cyclic operation, where the electron continuously shuttles between the left and right positions of Fig.~\ref{schematics}. There is considerable freedom in designing such cycle. However, we will first assume that the two shuttling processes between $x=\pm L/2$ are sufficiently fast to treat them as instantaneous quenches (in the spin degrees of freedom). This is not in contrast with the adiabatic assumption about orbital degrees of freedom, since typical orbital energies are much larger than the Zeeman splitting. A detailed discussion of shuttling with finite speed is given in Sec.~\ref{sec:initialization}. 

In summary, the mode of operation is a four-step cycle illustrated in Fig.~\ref{schematics}(b) and comprised by: (i) initialization period $t_l$ at $x=-L/2$, loading a single electron in the $\downarrow$ state; (ii) a fast shuttling process to the right operation point; (iii) a waiting period  $t_r$ at $x=L/2$, when the electron interacts with the nuclear spins allowing for flip-flop processes to occur; (iv) fast shuttling back to $x=-L/2$. Effectively, we treat the cycle as a two-step process with only (i) and (iii) and the period is $T \simeq t_l+t_r$. In the first part of each cycle we describe the evolution using:
\begin{equation}\label{ME_Lindblad_2a}
\dot{\rho}_{s} = 
-i[H_{z} ,\rho_{s}]+\Gamma_{l}(\mathcal{D}[d_\uparrow] + \mathcal{D}[d_\downarrow^{\dagger}])\rho_{s},
\end{equation}
where $H_z$ is the Zeeman Hamiltonian, defined by taking $A=0$ in Eq.~(\ref{H_qdot}), and the dissipator is of the Lindblad type, $\mathcal{D}[L]\rho_s=L\rho_s L^{\dagger}-\frac{1}{2}\{L^{\dagger}L,\rho_s \}$. Equation~(\ref{ME_Lindblad_2a}) is a standard master equation for a quantum dot in contact with an external reservoir (the left lead) where the chemical potential $\mu_l $ lies between the two Zeeman levels. $\rho_{s}$ is the full density matrix of the system, i.e., includes both the electronic and nuclear degrees of freedom, but the nuclear dynamics is trivial in this case.

For the second part of each cycle (e.g., $t_l< t< T$), the quantum dot center lies on the top of the nuclear-spin island and it is important to take into account the hyperfine interaction. As described in Appendix \ref{section_1a}, we perform a standard derivation by tracing out the lead degrees of freedom in the second-order Born-Markov approximation. After a rotating-wave approximation (RWA), we obtain:
\begin{equation}\label{ME_Lindblad_2b}
\dot{\rho}_{s} = 
-i[H_{0} ,\rho_{s}]  +\Gamma_{r}\sum_{\sigma}(\mathcal{D}[A_{\sigma+}] + \mathcal{D}[A_{\sigma-}^{\dagger}])\rho_{s},
\end{equation}
where we defined the Lindblad operators:
\begin{equation}\label{Asigma_pm}
A_{\sigma \pm} = d_\sigma  P_\pm ,
\end{equation}
with $P_\pm = \sum_{I,m}|\varphi_{I,m}^\pm \rangle\langle  \varphi_{I,m}^\pm |$ the projectors on the one-electron eigenspaces, see Eq.~(\ref{eigenstates}). For large Zeeman splitting (compared to the strength of hyperfine interaction), one has $ A_{\uparrow +}\simeq d_\uparrow $ and $ A_{\downarrow -}\simeq d_\downarrow $ while $ A_{\uparrow -},  A_{\downarrow +}\simeq 0$. However, in general it is important to take into account consistently the hyperfine interaction both in the Hamilonian and dissipative terms. As we will discuss in more detail in Sec.~\ref{sec:comparison}, the small difference between $d_\sigma$ and  $ A_{\sigma \pm}$ can have important effects on the long-time evolution.

\begin{figure}
\begin{centering}
\includegraphics[width=0.45\textwidth]{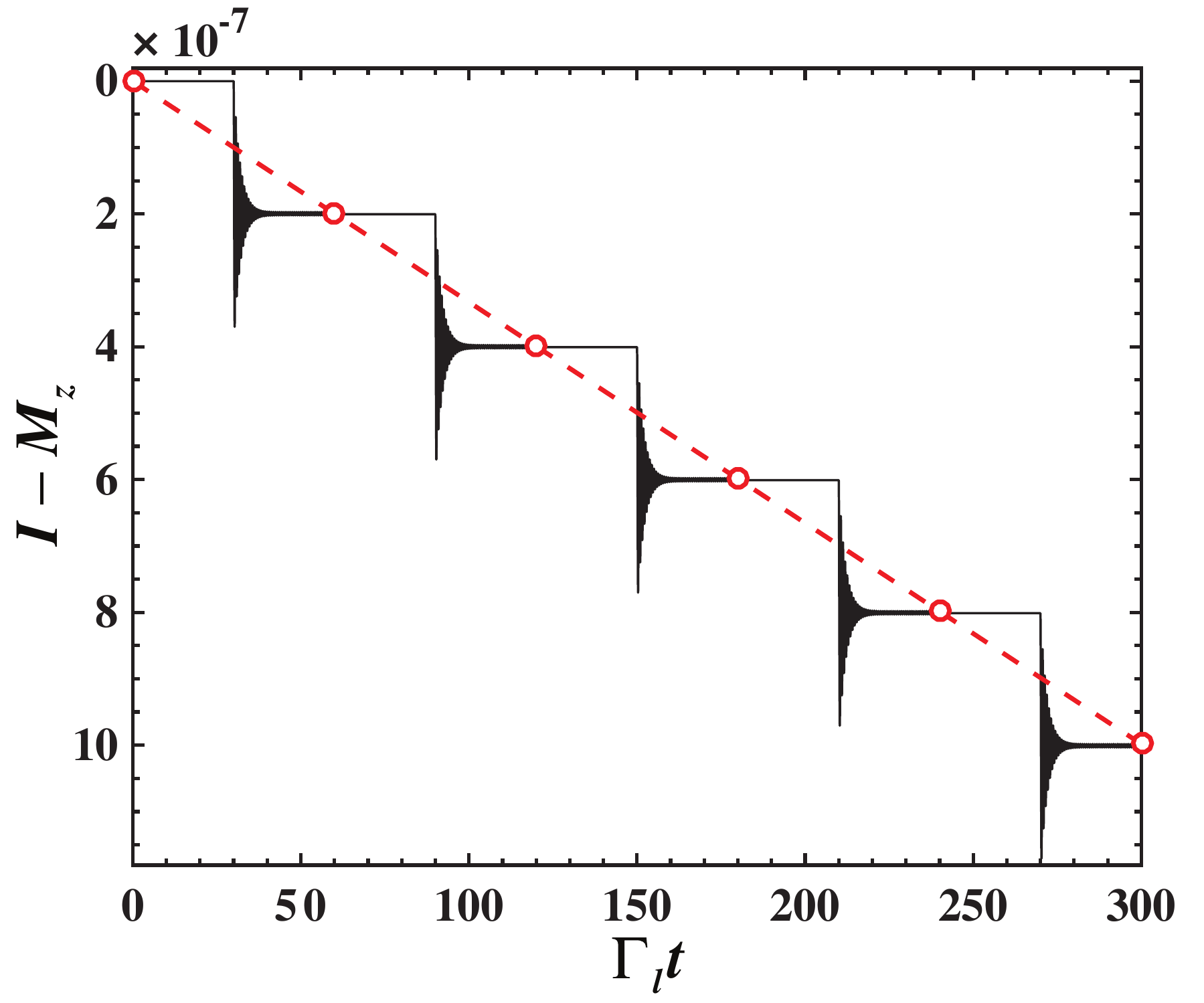}
\par\end{centering}
\caption{Full dynamics of the cyclic shuttling process, described by sudden
quenches between a right and left waiting periods, see Eqs.~(\ref{ME_Lindblad_2a}--\ref{ME_Lindblad_2b}).
Starting from $|\downarrow, N/2 \rangle$ (i.e., fully polarized nuclei), we plot the full time-dependence of the nuclear-spin polarization (solid black curve) as well as
its approximate stroboscopic evolution (red dashed line).
Parameters (setting $\omega_{0}=1$): $N=10$, $I=N/2$, $\eta=1\times10^{-4}$,
$\Gamma_{l}=\Gamma_{r}=0.1$, $\mu_{l}=\mu_{r}=\bar{\epsilon}=0$,
and $t_{l}=t_{r}=  30 ~\Gamma_{l,r}^{-1}$. }\label{shuttling_cycles}
\end{figure}

An example of the numerical results obtained in this manner is shown in Fig.~\ref{shuttling_cycles}, where the detailed evolution of the nuclear spin polarization  $M_{z}(t)\equiv\mathrm{Tr}\{I^{z}\rho_{s}(t)\}$ is plotted. The result is that the periodic shuttling leads to a systematic lowering of the nuclear spin polarization with each half-cycle. The physical mechanism behind this effect is directly related to the form of the eigenstates Eq.~(\ref{eigenstates}), which are superpositions of $|\downarrow,m \rangle$ and  $|\uparrow,m-1 \rangle$, i.e., they take into account the exchange of angular momentum between electron and nuclear spins induced by the hyperfine interaction. Since $|\varphi_{I,m}^\pm\rangle$ differ form the Zeeman eigenstates, a fast shuttling processes leads to a small probability of populating the high-energy eigenstate and allows the electron to tunnel out of the dot. Such processes are effectively associated with a flip-flop of electron and nuclear spins, thereby lowering $M_z$.

The full time evolution eventually leads to a full reversal of the nuclear spin polarization. However, the drop of $M_z$ at each cycle is small due to the small amplitude of flip-flop states in Eq.~(\ref{eigenstates}): $\beta_m \propto \eta$ gives a change in magnetization $\Delta M_z \propto \eta^2$ [see also the discussion in Sec.~\ref{sec:comparison}, and especially Eq.~(\ref{Mz_approx})]. Therefore, the superradiant-like enhancement appears after many cycles, which are numerically cumbersome to simulate. In the next section we develop an approximate stroboscopic treatment which is accurate (see dashed line in Fig.~\ref{shuttling_cycles}) and is able to describe the dynamics in a more efficient and physically transparent manner.

\subsection{Stroboscopic description}

If, as in Fig.~\ref{shuttling_cycles}, the waiting times $t_{r,l}$ are relatively long compared to $\Gamma_{l,r}^{-1}$, the system approaches a (temporary) stationary state before each quench. Under these conditions, it is possible to derive a simpler stroboscopic description of the long-time evolution. More specifically, the system after $n$ periods is described by:
\begin{equation}
\rho_s(nT) \simeq |\downarrow \rangle \langle \downarrow | \otimes \sum_m p_m(n)|I,m \rangle \langle I,m | ,
\end{equation}
and the nuclear-spin bath populations are determined by the discrete time evolution:
\begin{equation}\label{p_evolution}
{\bf p}(n) =  {\bf A}^n {\bf p}(0),
\end{equation}
where ${\bf p}(n)=(p_{-I}(n),p_{-I+1}(n),\ldots p_I(n))^T$ and the evolution matrix ${\bf A}$ is derived below.

To obtain ${\bf A}$, we first consider the electron prepared at the left position in the state $|\downarrow \rangle \otimes |I,m+1\rangle$. After the sudden quench to the right position, it is appropriate to use the eigenstates of Eq.~(\ref{eigenstates}), giving:
\begin{equation}\label{superpos_R}
|\downarrow , m+1\rangle = \alpha_{m} |\varphi_{I,m+1}^-\rangle + \beta_{m} |\varphi_{I,m}^+\rangle .
\end{equation}
This state constitutes the initial condition for Eq.~(\ref{ME_Lindblad_2b}) where, due to the RWA approximation, the coherence between $|\varphi_{I,m+1}^-\rangle$ and $|\varphi_{I,m}^+\rangle$ decays to zero without affecting the population dynamics. Thus, the stationary state is determined by rate equations alone. 

\begin{figure}
\begin{centering}
\includegraphics[width=0.4\textwidth]{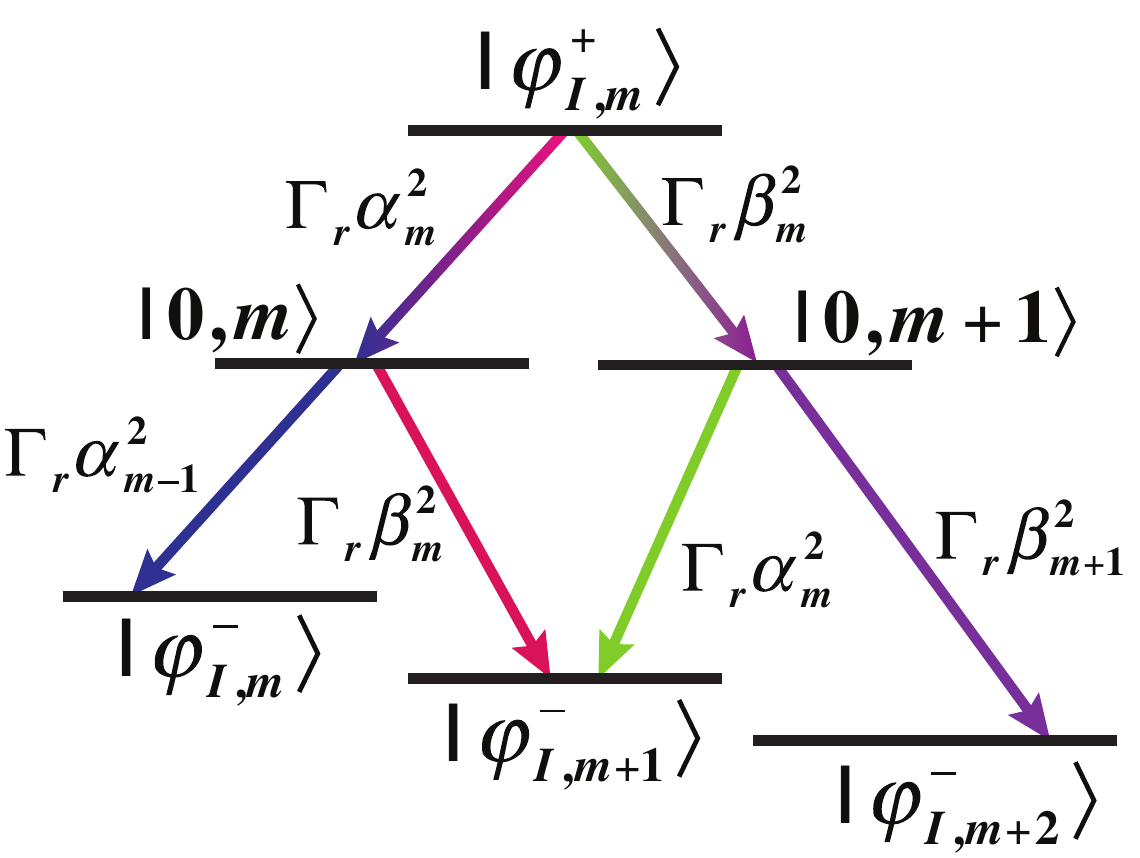}
\par\end{centering}
\caption{Branching processes for the state $|\varphi_{I,m}^+\rangle$. The rate of each process is indicated explicitly. The rate equations can be easily obtained from Eq.~(\ref{ME_Lindblad_2b}) and are  explicitly given in Eq.~(\ref{rate_eqs}), setting $\Gamma_l =0$. }\label{branching}
\end{figure}

While $|\varphi_{I,m+1}^-\rangle $ is already stationary, the high-energy state $|\varphi_{I,m}^+\rangle $ leads to the electron tunneling out of the quantum dot, followed by a process where the dot is re-occupied. The detailed branching processes for $|\varphi_{I,m}^+\rangle $, with the corresponding rates, are illustrated in Fig.~\ref{branching}. Taking them into account, it is seen that $|\downarrow , m+1\rangle$ evolves into a mixture of $|\varphi_{I,m}^-\rangle$, $|\varphi_{I,m+2}^-\rangle$, and $|\varphi_{I,m+1}^-\rangle$ and the final populations can be obtained as follows:
\begin{align}\label{R_matrix}
& R_{m,m+1} =  \frac{\beta^2_m  \alpha_m^2\alpha_{m-1}^2}{\alpha^2_{m-1}+\beta^2_{m}} , \quad  R_{m+2,m+1} =  \frac{\beta^4_m  \beta_{m+1}^2}{\alpha^2_{m}+\beta^2_{m+1}}  , \nonumber \\
& R_{m+1,m+1} = 1- R_{m,m+1} -R_{m+2,m+1} .
\end{align}

If we consider the reverse shuttling process, where the electron is prepared in a $|\varphi_{I,m}^{-}\rangle$ eigenstate and is quickly shuttled to the (left) nuclear-spin free region, the following sequence of tunneling events becomes possible for the component of $|\varphi_{I,m}^{-}\rangle$ in the excited state: $|\uparrow,m-1\rangle \to |0,m-1\rangle \to  |\downarrow,m-1\rangle$. It is quite clear that the final state will be a mixture of $|\downarrow,m\rangle$ and $|\downarrow,m-1\rangle$, and the populations are given by:
\begin{equation}\label{L_matrix}
L_{m,m} =  \alpha_{m-1}^2 , \quad  L_{m-1,m} =  \beta_{m-1}^2 .
\end{equation}
In summary, the effect of a full cycle at the left operating point is to induce transitions from an initial condition $|\downarrow, m+1\rangle$ to four final states: $|\downarrow, m-1\rangle, \ldots |\downarrow, m+2\rangle $ and the transition matrix $\bf A$ entering Eq.~(\ref{p_evolution}) is:
\begin{equation}\label{A_matrix}
{\bf A} = {\bf L}\, {\bf R},
\end{equation}
where the non-zero matrix elements of ${\bf L}, {\bf R}$ are given by Eqs.~(\ref{R_matrix}) and (\ref{L_matrix}), after a straightforward redefinition of the indexes (from $[-I,I]$ to $[1,2I+1]$).

\begin{figure}
\begin{centering}
\includegraphics[width=0.45\textwidth]{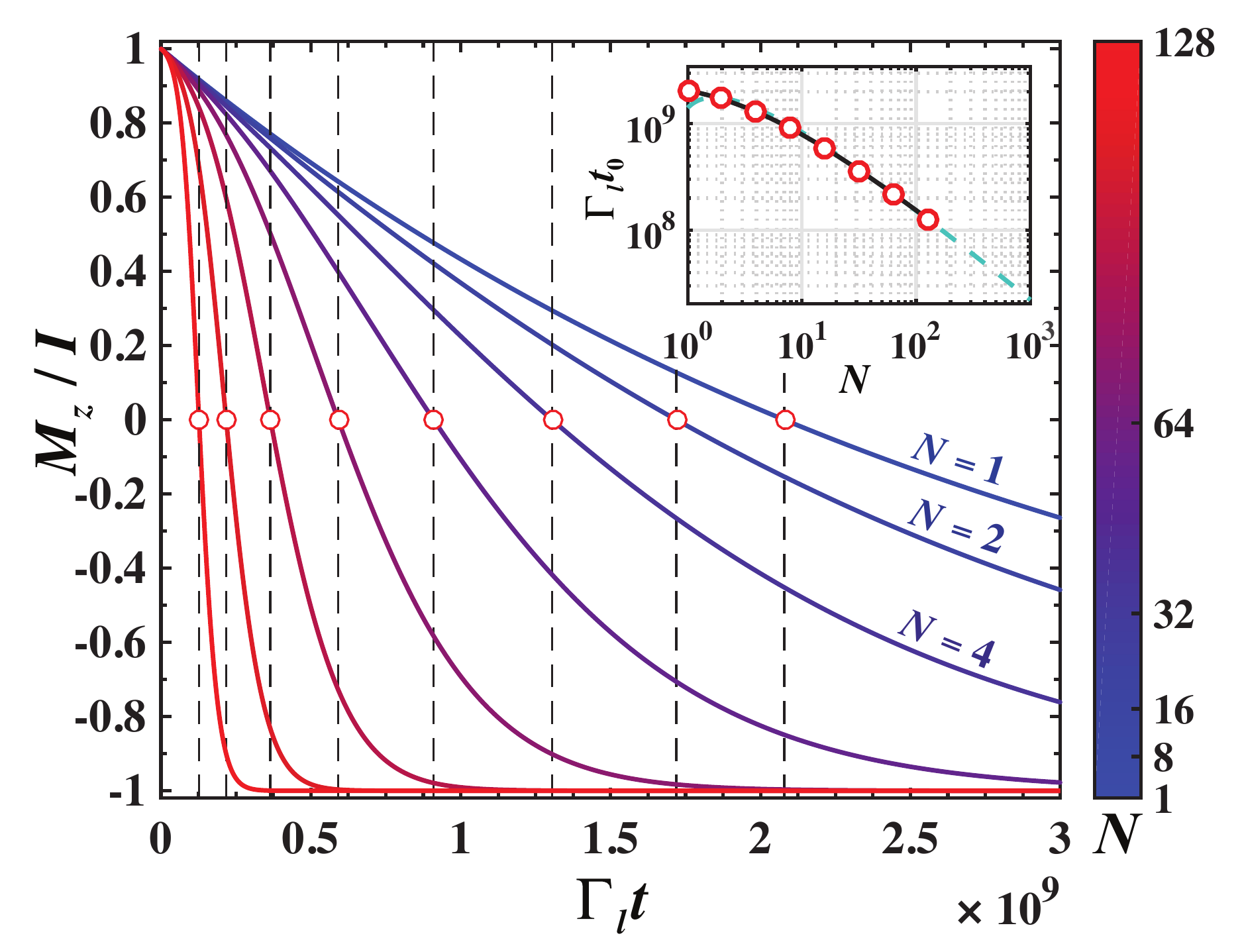}
\par\end{centering}
\caption{Time dependence of the nuclear spins polarization $M_{z}$ at different values of $N=1,2,4,\ldots 128$ (see color code). The vertical dashed
lines with red dots mark the times $t_{0}$ at which $M_{z}=0$. Inset: scaling of $t_{0}$ with respect to $N$, where the
blue dashed curve shows the theoretical prediction Eq.~(\ref{t0_SR}). Except $N$, other parameters are the same of 
Fig.~\ref{shuttling_cycles}. }\label{superradiance1}
\end{figure}

\begin{figure}
\begin{centering}
\includegraphics[width=0.45\textwidth]{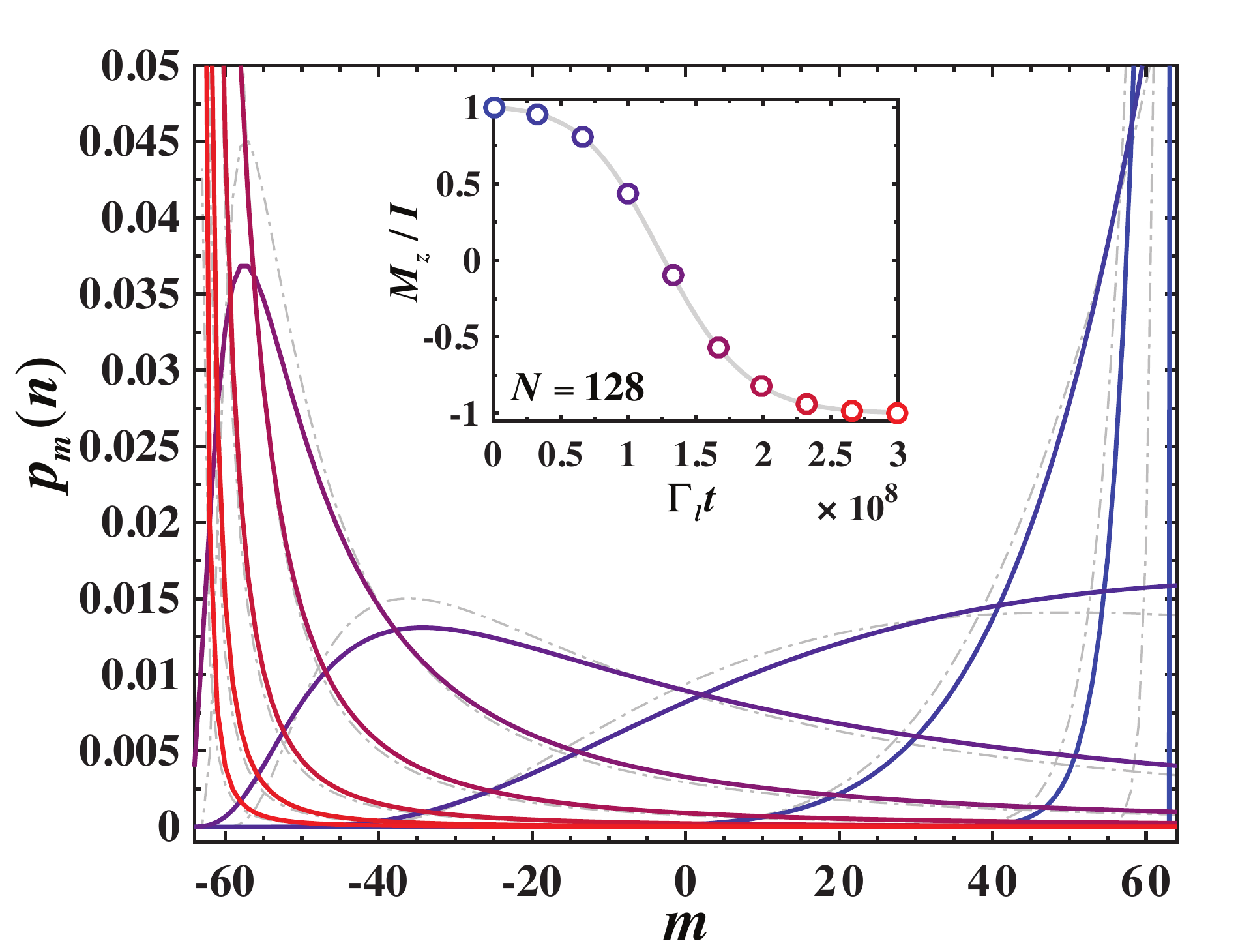}
\par\end{centering}
\caption{Time evolution of the distribution $p_{m}$ (solid lines),  for the $N=128$ case of Fig.~\ref{superradiance1}.
The time of each distribution is marked in the inset with a circle of the same color.
The gray dash-dotted curves show the superradiant approximation, Eq.~(\ref{pm_SR}).   }\label{superradiance2}
\end{figure}

As discussed in Fig.~\ref{shuttling_cycles}, we have checked that the stroboscopic description agrees well with the full time-dependence. We show in Fig.~\ref{superradiance1} the long-time evolution of the nuclear-spin polarization $M_z(t)$ at increasing values of $N=1,2,4,\ldots 128$ and illustrate in Fig.~\ref{superradiance2} the evolution of the full distribution function, $p_m$, in the case $N=128$. The behaviors of $M_z$ and $p_m$ are in good agreement with the features of optical superradiance. We see that the evolution time is reduced at larger values of $N$ and $p_m$ becomes a broad distribution with significant weight over all values of $m$. The large variance at intermediate times reflects the large shot-to-shot fluctuation typical of superradiance.\cite{1982_Phys_Rep_Gross,2018_Nat_Phys_Angerer}




\subsection{Connection to superradiance}

The previous stroboscopic description can be directly related to a standard description of Dicke superradiance.\cite{1982_Phys_Rep_Gross} To see this, we observe that the relative probabilities of the branching processes are controlled by the small parameter $\beta_m^2 \propto \eta^2$. The most likely event is $|\downarrow, m\rangle \to  |\downarrow, m\rangle$ which, however, does not affect $M_z$. Clearly, only the processes which change $m$ are interesting for the time evolution.

As inferred from Eqs.~(\ref{R_matrix}--\ref{L_matrix}), the most likely nuclear spin flip is $|\downarrow, m\rangle \to  |\downarrow, m-1\rangle$. More precisely,  the probability that such spin-flip occurs during the cycle time $T$ is given by:
\begin{align}\label{Gamma_SR_1}
T\Gamma_{m\to m-1}&=L_{m-1,m-1}R_{m-1,m}+L_{m-1,m}R_{m,m} \nonumber \\ 
& \simeq R_{m-1,m}+L_{m-1,m} ,
\end{align}
where in the second line we used $L_{m-1,m-1},R_{m,m}\simeq 1$ and only kept the terms of order $\eta^2$. The presence of two contributions corresponds to spin-flip events taking place either at the right or left contact.

The other types of spin-flips have smaller rates. For example, there is also process $|\downarrow, m+1\rangle \to  |\downarrow, m+2\rangle$ increasing the nuclear polarization but it has a much smaller rate, $\propto \beta^4_m  \beta_{m+1}^2 \propto \eta^6$. If we neglect them, we find that the nuclear system will slowly depolarize according to Eq.~(\ref{Gamma_SR_1}). More explicitly:
\begin{equation}\label{Gamma_SR_2}
\Gamma_{m \to m-1}   \simeq \frac{2\eta^2}{T} [I(I+1)-m(m-1)],
\end{equation}
where  we used $\beta_m \simeq \theta_m/2$ and approximated $\theta_m$ by the small $\eta$ limit of Eq.~(\ref{theta_m}). Such dependence of the depolarization rate on $m$ has the same form of the superradiant decay of an ensemble of $N$ atoms (if $I=N/2$).  We then can borrow the known results for the superradiant evolution. In particular, starting from a fully polarized state, $M_z(0)=N/2$, the depolarization time yielding $M_z(t_0)=0$ is given by:
\begin{equation}\label{t0_SR}
t_0 \simeq \frac{\ln( 1.6N)}{2N} T/\eta^2,
\end{equation}
which is in excellent agreement with the stroboscopic dynamics of Fig.~\ref{superradiance1} (see inset). The following approximate formula for the distribution $p_m$ can also be obtained, considering the limit of large $I=N/2$ and $t\gtrsim T/(2N\eta)$:\cite{1982_Phys_Rep_Gross}
\begin{equation}\label{pm_SR}
p_{m}(n)\simeq\left(\frac{2I}{I+m}\right)^{2}\exp\left[-2I\left(2\eta^{2}n+\frac{I-m}{I+m}e^{-4I\eta^{2}n}\right)\right].
\end{equation}
As shown in Fig.~\ref{superradiance2}, also for the full distribution we find good agreement with the superradiant evolution.

\section{Stochastic evolution and current noise}\label{stochastic_evolution}

\begin{figure}
\begin{centering}
\includegraphics[width=8.6cm]{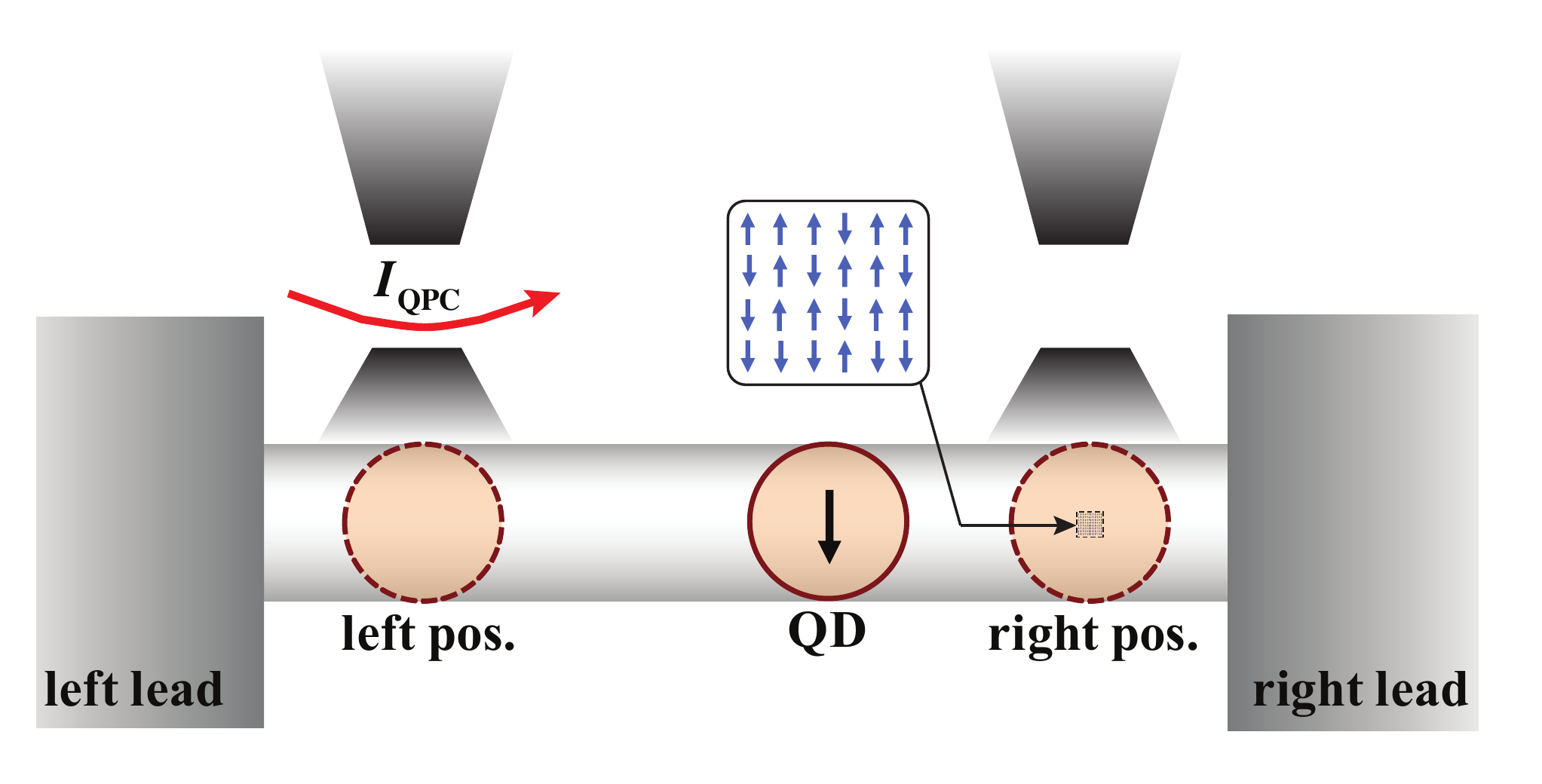}
\par\end{centering}
\caption{Schematics of the electron shuttle with two nearby charge sensors,
e.g., quantum point contacts (QPC). The sensors can perform non-demolition
measurements on the quantum dot occupation $n_{\mathrm{dot}}$ when
the dot is moved to the respective proximal positions. \label{FIG: QPC schematics}}
\end{figure}

While $\rho_s(t)$ gives the full ensemble-averaged evolution, the nuclear state is difficult to access directly. Therefore, the presence of nuclear-spin coherence should be inferred by charge or current measurements. An example is shown in Fig.~\ref{FIG: QPC schematics}, where we include two charge sensors at the left/right operation point to allow detecting individual tunneling events.\cite{}  In such a setup, a typical measurement would involve monitoring the quantum dot occupation and the superradiant-like dynamics will be reflected by the statistical properties of the tunnel events. Alternatively, it is also possible to measure the time-dependence of current noise through one of the contacts.

To address this type of evolution it is convenient to adopt a quantum-jump description of the master equation.\citep{1993_JOSAB_Molmer,1999_BOOK_Yamamoto} Following the standard prescription, the following collapse operators are introduced for Eq.~(\ref{ME_Lindblad_2a}):
\begin{equation}\label{Cleft}
C_{l,1}=\sqrt{\Gamma_{l}}d_{\uparrow},\text{ }C_{l,2}=\sqrt{\Gamma_{l}}d_{\downarrow}^{\dagger},
\end{equation}
and the collapse operators for Eq.~(\ref{ME_Lindblad_2b}) read:
\begin{align}
& C_{r,1}=\sqrt{\Gamma_{r}}A_{\uparrow+},& C_{r,2}=\sqrt{\Gamma_{r}}A_{\downarrow+}, \label{Cright1}\\ 
& C_{r,3}=\sqrt{\Gamma_{r}}A_{\uparrow-}^{\text{\dag}},& C_{r,4}=\sqrt{\Gamma_{r}}A_{\downarrow-}^{\text{\dag}}.   \label{Cright2}
\end{align}
In the periods between quantum jumps the electron and nuclear spins evolve according to an effective non-Hermitian Hamiltonian, $H_{z}-i/2\sum_{m}C_{l,m}^{\dagger}C_{l,m}$ or $H_{0}-i/2\sum_{m}C_{r,m}^{\dagger}C_{r,m}$ depending on the quantum dot's position. Since the jump operators in Eqs.~(\ref{Cleft}--\ref{Cright2}) correspond to projective measurements induced by the coupling with the left and right leads, they provide a direct connection between individual trajectories and the signal of charge sensors monitoring the quantum dot.

\begin{figure}
\begin{centering}
\includegraphics[width=8.6cm]{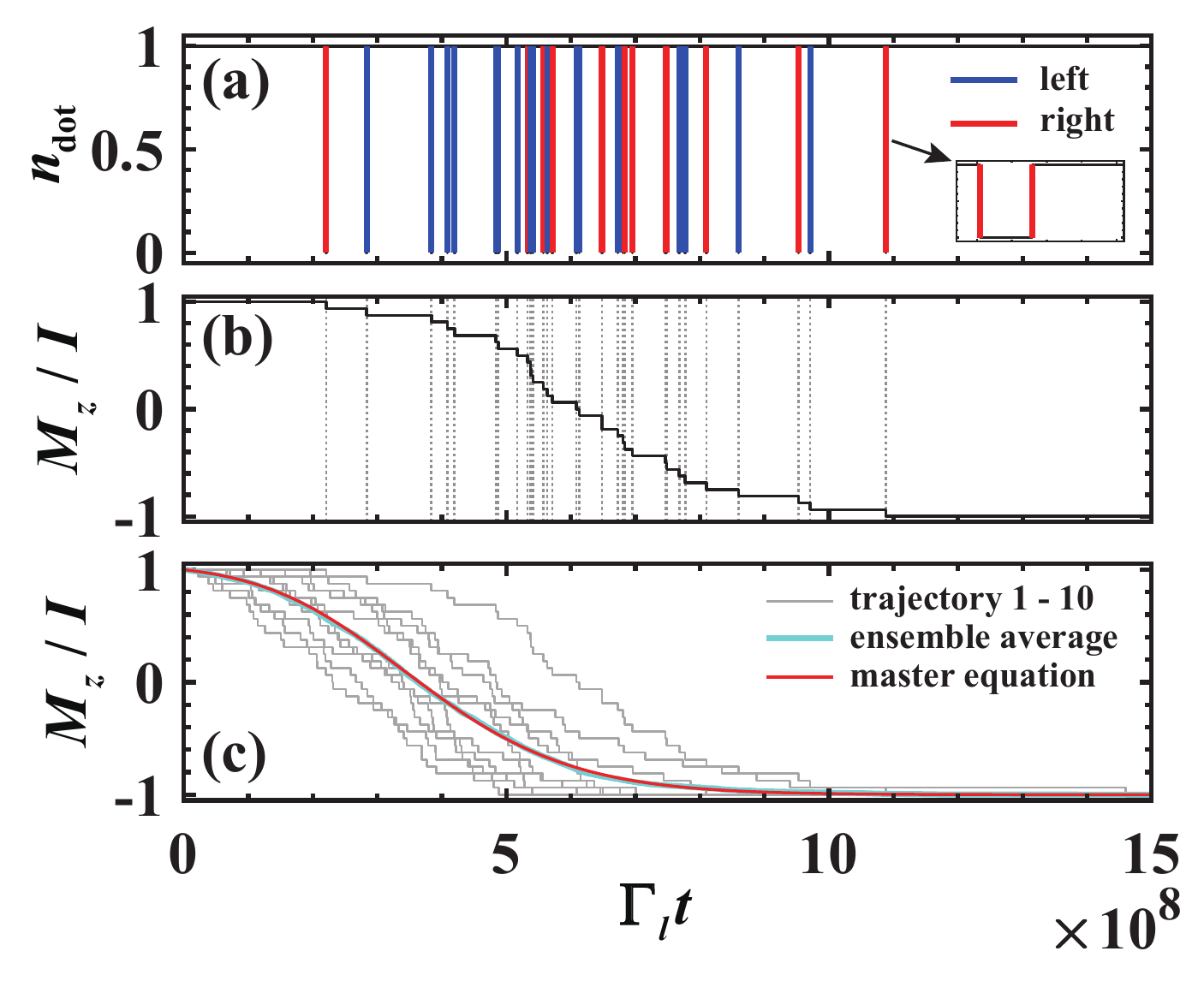}
\par\end{centering}
\caption{Charge sensing and nuclear spins polarization dynamics from the
Monte Carlo wave-function (MCWF) simulation. (a) Quantum dot occupation $n_\mathrm{dot}$ 
as a function of time for a representative MCWF trajectory. The blue (red) color marks tunneling events that happen when the dot is at the left (right) operating point. (b) Nuclear spins polarization $M_z/I$ (black solid) as a function of time, for the same trajectory of panel (a). The vertical dashed lines highlight the correspondence between jumps in polarization and tunneling events. (c) $M_z$ from an ensemble average over 100 MCWF trajectories (thick blue). The red thin curve is obtained from solving the stroboscopic evolution Eq.~(\ref{p_evolution}) with an initial distribution $p_m(0)=\delta_{m,I}$. The light gray curves shows the $M_z$ dynamics from 10 MCWF trajectories from the ensemble. Parameters used in the calculations (in unit of $\omega_{0}$): $N=32$, $I=N/2$, $\eta=10^{-4}$, $\Gamma_{l}=\Gamma_{r}=0.1$, and $t_{l}=t_{r}=300$.\label{FIG: MCWF counting statistics}}
\end{figure}

Figure \ref{FIG: MCWF counting statistics} illustrates a typical trajectory from the Monte Carlo wave-function simulation. We show in panel (a) the evolution of the quantum dot's occupation, characterized by a series of tunneling events where the electron jumps to the right/left contact and is immediately reloaded from it (see the inset). An important feature is the visible change in  frequency of tunneling events, which are much more rare at the beginning and the end of time evolution. The increase of frequency at intermediate times (despite the smaller number of nuclear spins which can be flipped) reflects the enhancement of tunnel rate induced by the nuclear coherence. A second important feature, illustrated in panel (b), is the direct correspondence of tunnel events to the quantum jumps in the nuclear-spin polarization. A change $|\Delta M_z| \simeq 1$ is associated with tunnel events occurring at both (left/right) contacts. Therefore, one can rely on charge measurements to monitor the nuclear-spin polarization.  Finally, we show in panel (c) that the ensemble-averaged nuclear-spin polarization coincides with the master equation treatment.

\begin{figure}
\begin{centering}
\includegraphics[width=8.6cm]{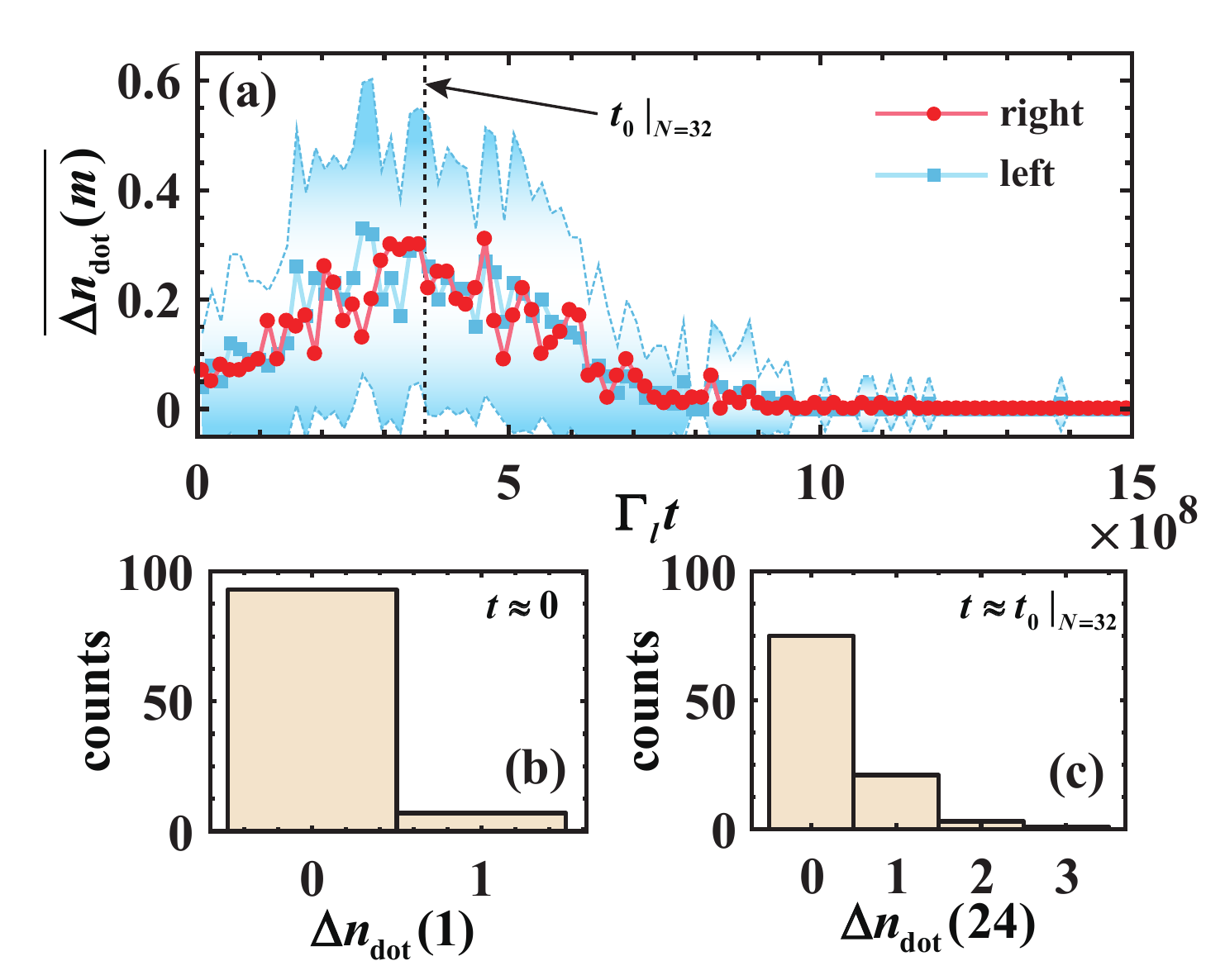}
\par\end{centering}
\caption{Statistics of $\Delta n_{\mathrm{dot}}(m)$ from the 100 MCWF trajectories of Fig.~\ref{FIG: MCWF counting statistics}. We consider a coarse-grained evolution with 100 intervals ($m=1,2, \ldots 100$). (a) Evolution of the average value, defined in Eq.~(\ref{dn_av}) and resolved between the left (blue squares) and right (red dots) contacts. The blue region indicates the fluctuations of $\Delta n_{\mathrm{dot}}(m)$ at the left contact, with the upper and lower bounds (dashed lines) given by $\overline{\Delta n_{\mathrm{dot}}(m)} \pm \sigma_{\rm dot}(m)/2 $ [see Eq.~(\ref{dn_sigma})].   Panels (b) and (c) show the distribution of $\Delta n_{\rm dot}(m)$ at the initial and intermediate stage evolution, respectively. The time assumed in panel (c),  is marked in panel (a) by a vertical dashed line. The two lower panels refer to the right contact, while the histograms of the left contact are almost identical . }\label{Currents_fluctuations}
\end{figure}

To quantify more precisely the occurrence of tunnel events, we consider a coarse-grained evolution over larger time intervals $\Delta t \gg T$, i.e., spanning many shuttling cycles. Since a trajectory $k$ (with $k=1,2,\ldots N_{\rm traj}$) is characterized by a series of random times $t_j^{(k)}$ ($j=1,2,\ldots$) at which the electron tunnels out of the quantum dot, we introduce $\Delta n_{\mathrm{dot}}^{(k)}(m) $ as follows:
\begin{equation}
\Delta n_{\rm dot}^{(k)}(m) = \int_{(m-1)\Delta t}^{m\Delta t}dt  \sum_{j} \delta(t-t_j^{(k)}),
\end{equation} 
which counts the number of narrow spikes in $n_{\rm dot}$ (see Fig.~\ref{FIG: MCWF counting statistics}) within the  $m$-th time interval. Operationally, the $t_j^{(k)}$s are detected from signal blips at the charge sensors. The average number during such $m$-th sub-period is:
\begin{equation}\label{dn_av}
\overline{\Delta n_{\mathrm{dot}}(m)}=\frac{1}{N_{\mathrm{traj}}}\sum_{k=1}^{N_{\mathrm{traj}}}\Delta n_{\mathrm{dot}}^{(k)}(m),
\end{equation}
and the fluctuation is given by
\begin{equation}\label{dn_sigma}
\sigma^2_{\rm dot}(m)=\overline{\Delta n_{\mathrm{dot}}(m)^{2}}-\overline{\Delta n_{\mathrm{dot}}(m)}^{2}.
\end{equation}

The evolution of these quantities with time, obtained numerically from a MCWF simulations of 100 trajectories, is shown in Fig.~\ref{Currents_fluctuations}(a). For each sub-interval, the distribution of tunneling events can also be extracted by direct histogram, with two examples shown in panels (b) and (c). Since we are dealing with a transient process, the form of the distribution evolves in time and, compared to the initial stage, develops an elongated tail around $t \sim t_0$ [see Eq.~(\ref{t0_SR})]. This dependence leads to the maximum in $\overline{\Delta n_{\mathrm{dot}}(m)}$ observed in panel (a). The increased frequency of tunnel events is also accompanied by stronger fluctuations in $\Delta n_{\mathrm{dot}}(m)$, reflecting the broad superradiant-like statistical distribution discussed in Fig.~\ref{superradiance2}. 

An interesting observation from Fig.~\ref{Currents_fluctuations} is that the behavior of the right and left contacts is essentially equivalent. Finally, we note that a detailed monitoring tunnel events might not be necessary. At variance with previous proposals\cite{2004_JPSJ_Eto, 2015_PRB_Stefano, 2012_PRB_Schuetz} here we do not apply a bias and there is zero average current flowing through the device [see, e.g., Fig.~\ref{Currents_fluctuations}(a), displaying a balanced number of tunnel in/out events at each contact]. Nevertheless, the evolution of $\overline{\Delta n_{\mathrm{dot}}(m)}$ reflects enhanced current fluctuations at intermediate times $t\sim t_0$. Therefore, an analysis of the time-dependent current noise at either one of the contacts should be able to reveal the coherent enhancement of tunnel rates induced by nuclear spins.

\section{Non-adiabatic shuttling process}\label{sec:initialization}

We now take a closer look at the shuttling process, and  discuss the regime of validity of treating it as an ideal quench. Clearly, this approximation is only appropriate below a certain shuttling time $t_{\rm f}$ and this timescale is critical for the superradiant-like dynamics: if the transfer from left to right is too slow, an initial $|\downarrow \rangle$ electron will evolve adiabatically into an eigenstate of the hyperfine Hamiltonian, and tunneling cannot take place. It is the purpose of this section to estimate how fast the shuttling time $t_{\rm f}$ should be.

\begin{figure}
\begin{centering}
\includegraphics[width=0.45\textwidth]{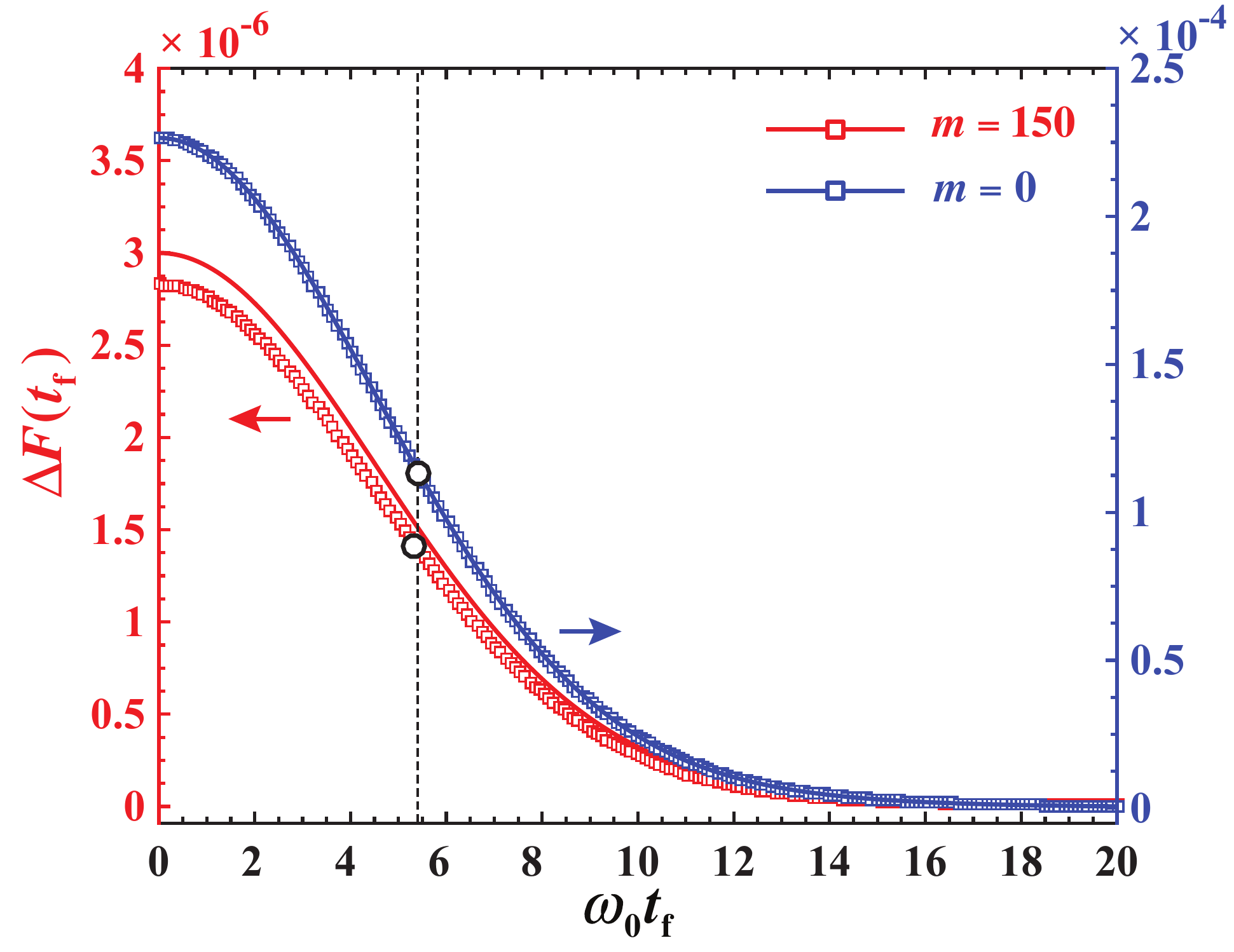}
\par\end{centering}
\caption{Probability of finding the quantum dot in the excited state at the end of the shuttling process. Here, the upper (blue) and lower (red) curves are for $m=0$ and 150, respectively (with $I=N/2=150$). Numerical results (squares) are compared to the approximate Eq.~(\ref{Fm_approx}) (solid lines). The times $t_*$ at which the probability dropped to one half of the initial value are marked by white dots. The vertical line is our estimate of $t_*$, Eq.~(\ref{tstar_hf}). We used $\Delta x =L/3$, and $ \eta_{\rm f} ^{-1}= 1 \times 10^4$. }\label{shuttling_fig}
\end{figure}

To this end, we rewrite $H_0(t)$ in the subspace spanned by $|\uparrow,m-1\rangle$ and $|\downarrow,m\rangle$. This basis defines pseudo-Pauli operators $\tilde{\sigma}_{i}$, e.g., $\tilde{\sigma}_z = |\uparrow,m-1\rangle \langle \uparrow,m-1 | - |\downarrow,m\rangle \langle \downarrow,m|$. Omitting a time-dependent constant we arrive at:
\begin{equation}\label{H0_truncated}
H_{0}(t)=\eta_t\omega_{0} \sqrt{I(I+1)-m(m-1)}\tilde{\sigma}_{x}+\frac{\omega_m(t)}{2}\tilde{\sigma}_{z},
\end{equation}
where:
\begin{equation}
\omega_{m}(t)=\omega_{0}[1+\eta_t (2m-1)],
\end{equation}
and $\eta_t= A(t)/(2N_d\omega_0)$. Assuming that the shuttling takes place with constant velocity, Eq.~(\ref{Ax}) gives:
\begin{equation}\label{eta_t}
\eta_t = \eta_{\rm f} e^{-(t/t_{\rm f}-1)^2 L^2/\Delta x^2},
\end{equation}
and we initialize the quantum dot in the $t=0$ ground state $| \varphi_{I,m}^{-}\rangle \simeq |\downarrow,m \rangle $. After evolving this state according to $H_0(t)$, we compute the probability $ \Delta F_m(t_{\rm f}) $ of finding the quantum dot in the excited state $| \varphi_{I,m-1}^{+}\rangle$ (with $\eta_{t=t_{\rm f}} $ much larger than $\eta_{t=0}$). A numerical evaluation of Eq.~(\ref{H0_truncated}) is shown in Fig.~\ref{shuttling_fig}, as function of $t_{\rm f}$.  The largest probability is obtained for an instantaneous transfer (the quench dynamics of previous sections), giving:
\begin{equation}\label{delta_m}
\Delta F_m(0) \simeq  \eta^2_{\rm f} \left[ I(I+1)-m(m-1) \right],
\end{equation}
which is in direct correspondence to Eq.~(\ref{Gamma_SR_2}). 

Although the initial value Eq.~(\ref{delta_m}) has a strong dependence on $m$, reflecting the enhancement of spin-flip probability around $m\sim0$, we see in Fig.~\ref{shuttling_fig} that the subsequent decay occurs on a timescale which is only weakly dependent on $m$. To gain insight into this time-dependence we apply ordinary time-dependent perturbation theory, which is justified by the small value of $\eta_t$. We find:
\begin{equation}\label{Fm_approx}
\Delta F_m(t_{\rm f}) \simeq  \Delta F_m(0) g\left(\frac{\omega_0 t_{\rm f}}{2}\frac{\Delta x}{L}\right),
\end{equation}
where $g(x)= |1-i\sqrt{\pi} x \exp[-x^2]\mathrm{erfc}(ix)|^2 $, with $\mathrm{erfc}(x)=1-{\rm erf}(x) $ the complementary error function.\cite{1970_BOOK_Abramowitz} To derive this expression we supposed $L/\Delta x \gg 1$. However, as shown in Fig.~\ref{shuttling_fig}, we find that Eq.~(\ref{Fm_approx}) becomes accurate already at moderate values $L/\Delta x \sim 2-3$. 

Importantly, $g(x)$ is independent of $m$ and allows us to identify the relevant timescale as $\omega_0^{-1}L/\Delta x$. For example, setting $\Delta F(t_{\rm f}) \ll \Delta F(0)/2$ one gets:
\begin{equation}\label{tstar_hf}
t_{\rm f} \gg t_*\approx 1.8   \omega_0^{-1} \frac{L}{\Delta x}.
\end{equation}
The physical interpretation of Eq.~(\ref{tstar_hf}) is rather transparent, after noticing that hyperfine interaction is exponentially suppressed in the first part of the shuttling process. A significant change of the Hamiltonian happens on a distance $\sim \Delta x$ rather than $L$, which effectively shortens the transfer time by a factor $\sim \Delta x/L$. Therefore, the energy is undetermined by an amount $\delta E \sim (t_{\rm f}\Delta x/L)^{-1}$. If this energy scale is much smaller than the gap $\omega_0$ between $\pm$ branches, the probability of being in the excited states is negligible [in agreement with Eq.~(\ref{tstar_hf})].

\begin{figure}
\begin{centering}
\includegraphics[width=0.45\textwidth]{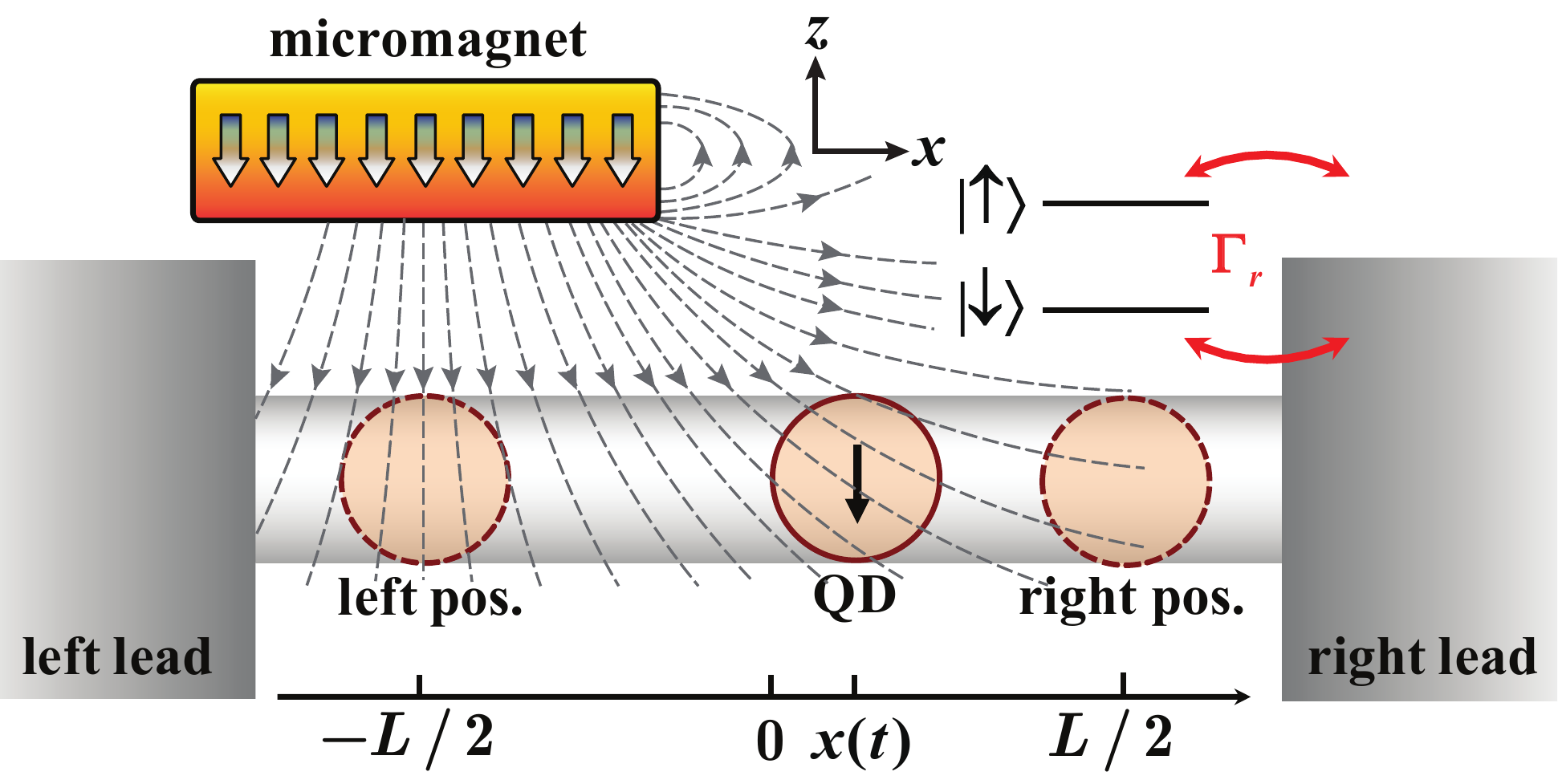}
\par\end{centering}
\caption{Schematics of an alternative shuttling setup. Here there are no nuclear spins but the electron shuttles through the inhomogeneous field of a nearby micromanget. In addition, a large homogeneous field is applied along $z$.  }\label{micromagnet_fig}
\end{figure}

In closing this section, we note that the problem described by Eq.~(\ref{H0_truncated}) is very similar to the scenario illustrated in  Fig.~\ref{micromagnet_fig}, where the shuttling takes place in the presence of a nonuniform magnetic field generated by a micromagnet.\cite{2006_PRL_Tokura,2008_Nat_Phys_Pioro,2014_PRB_Chesi} The main difference is that in that case the relatively small variation of the magnetic field can be taken as approximately linear (supposing a shuttling process with constant velocity $L/t_{\rm f}$). If the time-dependence is of the type:
\begin{equation}\label{B_t}
H_{B} =  \frac{\omega_0}{2} \left [  \delta_\perp \frac{t}{t_{\rm f}} \sigma_x   +\left(1+\delta_\parallel \frac{t}{t_{\rm f}}\right) \sigma_z  \right],
\end{equation}
where $| \delta_{\perp,\parallel}| \ll 1$, the probability of being in the excited state at the end of the transfer process (starting from $|\downarrow \rangle$) can be computed as follows:
\begin{equation}\label{F_micromagnet}
\Delta F(t_{\mathrm{f}})\simeq \frac{\delta_\perp^2}{4}\frac{\sin^2\left(\omega_0 t_{\rm f}/2\right)}{(\omega_0 t_{\rm f})^2/4},
\end{equation}
giving the characteristic timescale:
\begin{equation}\label{tstar_B}
t_* \sim 2.8 \omega_0^{-1}.
\end{equation}
We see that also in this case $t_*$ is determined by the Zeeman splitting. For $t_{\rm f} \gg t_*$, the shuttling process is slow and allows the spin to adjust to the instantaneous field. On the other hand, if $t_{\rm f} \lesssim t_*$, the electron will have a probability $\gtrsim \frac18 \delta_\perp^2$ to be excited at the end of the transfer process and, with a bias configuration like in Fig.~\ref{micromagnet_fig}, can tunnel out of the quantum dot. 

In summary, we find that the typical timescale of shuttling processes inducing an electron spin-flip is given by the inverse Zeeman energy: both for the nuclear-spin island and the micromagnet a shuttling time of order $\omega_0^{-1}$ has an effect similar to the instantaneous transfer [see Eq.~(\ref{tstar_hf}) and (\ref{tstar_B}), respectively].

\section{Shuttling vs. stationary configurations}\label{sec:comparison}

The superradiant-like dynamics of nuclear spins in single quantum dots was discussed before in Refs.~\onlinecite{2012_PRB_Schuetz} and \onlinecite{2015_PRB_Stefano} where, however, stationary configurations were considered (with no shuttling). We would like to highlight in this section what are the main differences and potential advantages of the shuttling configuration.

With respect to the quantum-dot spin valve proposed in Ref.~\onlinecite{2015_PRB_Stefano}, an advantage of the present setup is that it does not require the fabrication of ferromagnetic leads.\cite{2000_PRB_Schmidt,2003_JPD_Jasen,2006_PRL_Zutic,2010_APL_Aurich,2011_JAP_Tarun}   Instead, the scheme analyzed in Ref.~\onlinecite{2012_PRB_Schuetz} considers an ordinary quantum dot in the weak tunneling regime, with a simple level structure and normal leads. That proposal represents an attractive option, but we find that in a revised theoretical description the superradiant-like transport features disappear, suggesting that a non-adiabatic process analogous to the fast shuttling is necessary.

\begin{figure}
\begin{centering}
\includegraphics[width=0.45\textwidth]{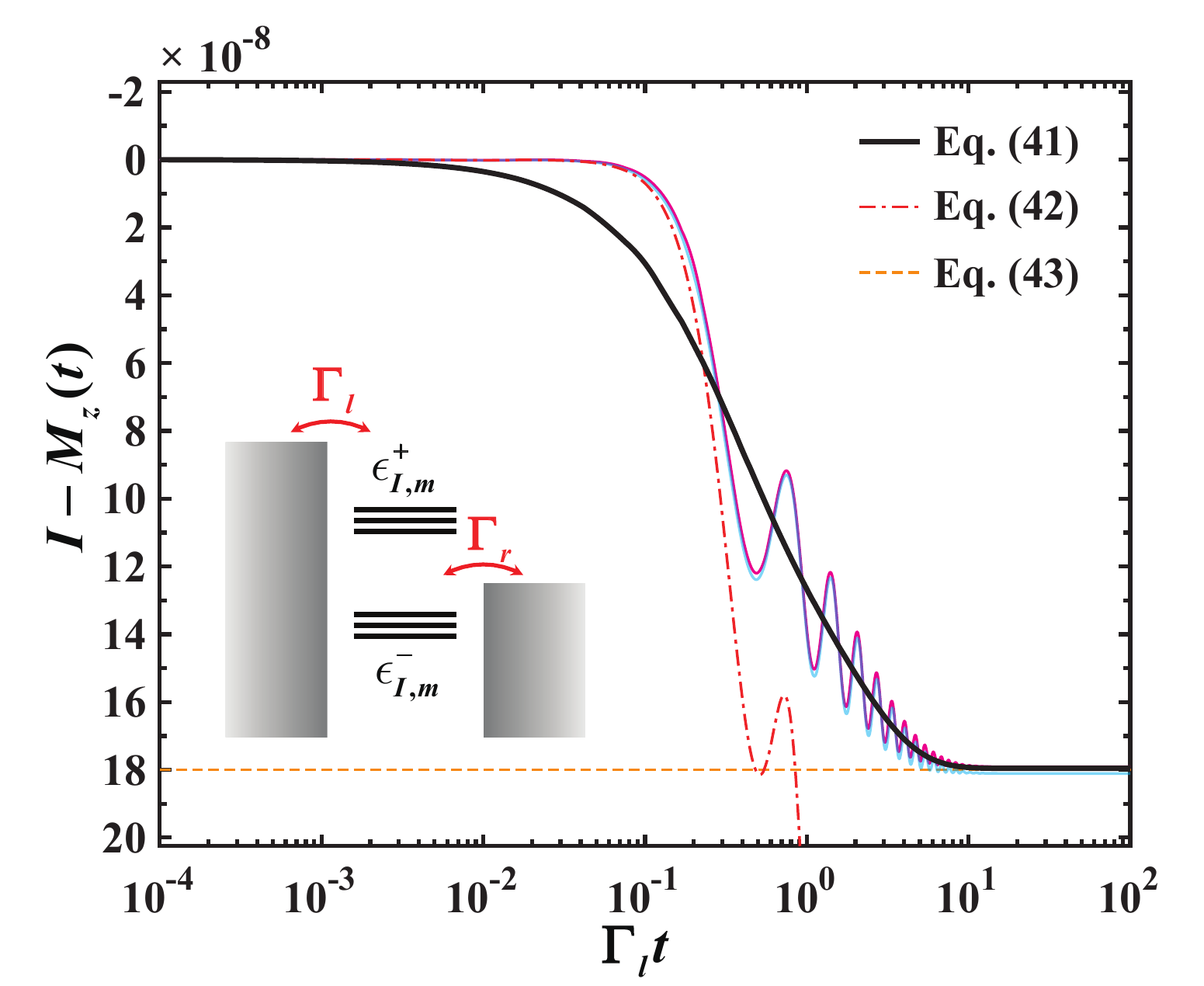}\label{compare}
\par\end{centering}
\caption{Nuclear spins polarization versus time in a static quantum dot configuration (see inset). The black solid curve shows $M_z$ from solving the Lindblad master equation, Eq.~(\ref{ME_Lindblad}). The thin blue curve is from Eq.~(\ref{m_eq_1}), i.e., without performing the RWA and taking into account the Lamb shift  (the thin magenta curve shows the small effect of setting $\Delta_{LS}=0$). The red dash-dotted curve is the more approximate evolution Eq.~(\ref{ME_superrad}). Parameters used in the calculations (in unit of $\omega_0$): $N=12$, $I=N/2$, $\eta=10^{-4}$, $\Gamma_l=\Gamma_r=0.1$, $\mu_l=2$, $\mu_r=0$, and $\Lambda=10^3$ [cf. Eq.~(\ref{Delta_LS})].}
\end{figure}

To clarify this point, we consider in detail the transport model illustrated in the inset of Fig.~\ref{compare}. Since the position of the dot is kept fixed, the Hamiltonian is simply given by Eq.~(\ref{H}), with time-independent tunneling amplitudes and hyperfine coupling strength. An external bias is applied, with the $\epsilon^{+}_{I,m} $ levels lying in the transport window. The main simplifications with respect to Ref.~\onlinecite{2012_PRB_Schuetz} are that we restrict ourselves to a uniform hyperfine coupling and large Zeeman splitting, such that we can avoid including a dynamical compensation of the longitudinal Overhauser field (along $z$).  We derive the master equation as in Sec.~\ref{sec:evolution} (and Appendix \ref{section_1a}), obtaining:
\begin{align}\label{ME_Lindblad}
\dot{\rho}_{s} =  -i[H_{0} & ,\rho_{s}]+\sum_{\sigma}\Big(\Gamma_{r}\mathcal{D}[A_{\sigma+}]\rho_{s} \nonumber \\
&+\Gamma_{l}\mathcal{D}[A_{\sigma+}^{\dagger}]\rho_{s} 
+(\Gamma_{r}+\Gamma_{l})\mathcal{D}[A_{\sigma-}^{\dagger}]\rho_{s}\Big),
\end{align}
where the Lindblad operators are given in Eq.~(\ref{Asigma_pm}).  A  numerical example of the typical nuclear polarization dynamics (starting with an empty quantum dot, $|0\rangle\otimes|I,I\rangle$) is presented in Fig.~\ref{compare}. 

The most remarkable feature of of Fig.~\ref{compare}  is the small change in $M_z$, which is in contrast to the full polarization reversal predicted for superradiant-like dynamics.  The stationary state is determined by the special form of the Lindblad operators $A_{\sigma \pm }$, which involve projectors on the $\pm$ bands. Therefore, the $|\varphi_{I,m}^{-}\rangle$ eigenstates are stationary solutions of the master equation, inhibiting further dynamics.  According to Eq.~(\ref{ME_Lindblad}), the nuclear spin bath is unable to remove the Coulomb blockade and, once the quantum dot is occupied in the $-$ band, there are no further spin-flips affecting the nuclear-spin polarization.  

Based on Eq.~(\ref{ME_Lindblad}), we can give an approximate expression of the small polarization loss from a rate equation analysis, using the fact that $\eta$ is small. This approach is described in detail in Appendix \ref{sec_rates} and here we only cite the final result for the stationary value. For  $I=N/2$:
\begin{equation}\label{Mz_approx}
M_z (t\to \infty)\simeq \frac{N}{2}\left(1-2\eta^2\frac{2\Gamma_{l}+\Gamma_{r}}{\Gamma_{l}+\Gamma_{r}}\right),
\end{equation}
showing that the depolarization is indeed small when $\eta \ll 1$. We have also extended the above analysis by evaluating the higher order corrections to $M_z$, see Eq.~(\ref{Mz_perturbative}). 

To check that the behavior is not an artifact of the RWA between the $\pm$ bands, we have also integrated numerically Eq.~(\ref{m_eq_1}), which only relies on the second-order Born-Markov approximation (justified in the weak-tunneling regime $\Gamma_{l,r} \ll \omega_0$). As expected, this treatment displays a short-time oscillatory dynamics absent under RWA. Otherwise, as shown in Fig.~\ref{compare}, the two approaches agree on the general features of the time-dependence and, most importantly, on the small change of the spin polarization. 

On the other hand, the long-time behavior dramatically changes by neglecting the hyperfine interaction in the dissipator, which leads to a superradiant-like master equation:\cite{2012_PRB_Schuetz,2015_PRB_Stefano}
\begin{align}\label{ME_superrad}
\dot{\rho}_{s} \simeq  -i[H_{0}  ,\rho_{s}]& +\Gamma_{r}\mathcal{D}[d_{\uparrow}]\rho_{s}  +\Gamma_{l}\mathcal{D}[d_{\uparrow}^{\dagger}]\rho_{s} \nonumber \\
&+(\Gamma_{r}+\Gamma_{l})\mathcal{D}[d_{\downarrow}^{\dagger}]\rho_{s}.
\end{align}
A numerical solution of Eq.~(\ref{ME_superrad}) is shown in Fig.~\ref{compare}, where the saturation of $M_z$ is not observed in this case. However, we stress that Eq.~(\ref{ME_superrad}) involves an additional approximation with respect to Eq.~(\ref{ME_Lindblad}).

To understand the disagreement between the two master equations we note that Eq.~(\ref{ME_superrad}) can be justified at any given timescale when the hyperfine coupling $A$ is  sufficiently small. In that limit, indeed $A_{\uparrow,+}\simeq d_\uparrow$ and $A_{\downarrow,-}\simeq d_\downarrow$ [see after Eq.~(\ref{Asigma_pm})]. However, when $A \to 0$, it also happens that the rate of flip-flop processes decreases quickly, being proportional to $A^2$. Correspondingly, the predicted timescale of the superradiant-like evolution grows like $\propto A^{-2}$. On this diverging timescale, the small difference in propagators between Eqs.~(\ref{ME_superrad}) and (\ref{ME_Lindblad}) leads to important deviations. From Fig.~\ref{compare} we conclude that the threshold time for Eq.~(\ref{ME_superrad})  [i.e., the time after which it becomes inaccurate]  must be shorter than the predicted superradiant-like timescale.

In the light of these discussions one can appreciate better the crucial role played in our proposal by the non-adiabatic shuttling processes, which allows to overcome the blockaded regime and induce the desired superradiant-like evolution.

\section{Conclusion}

In this work we have analyzed the combined electron-nuclear spin dynamics in an electron shuttling device with a strongly inhomogeneous distribution of nuclear spins. We have shown that, under suitable conditions, it is possible to generate quantum coherence in the nuclear spin system through collective electron-nuclear flip-flop processes. Similarly to Refs.~\onlinecite{2004_JPSJ_Eto,2012_PRB_Schuetz,2015_PRB_Stefano}, the nuclear-spin dynamics follows a superradiant-like evolution reflected in charge transport, i.e., leading to a large enhancement of the effective tunneling rates. 

One important condition for the superradiant-like dynamics to take place is the non-adiabaticity of the shuttling process. This requirement is related to potential difficulties in removing the Coulomb blockade in static devices in the weak-tunneling regime. Taking advantage of a fast shuttling dynamcs, our proposal would allow the superradiant-like evolution to take place without relying on ferromagnetic leads or multi-dot setups.\cite{2004_JPSJ_Eto,2015_PRB_Stefano}

Despite these differences, the basic mechanisms at the core of the superradiant-like evolution is the same of previous proposals.\cite{2004_JPSJ_Eto,2012_PRB_Schuetz,2015_PRB_Stefano} Therefore, similar considerations about timescales and regimes of validity apply. In particular, the effects of inhomogeneous hyperfine coupling, imperfect initial polarization, and nuclear-spin decoherence were already analyzed in Refs.~\onlinecite{2004_JPSJ_Eto,2012_PRB_Schuetz,2015_PRB_Stefano} and we expect minor differences in our case. 

Here we only point out that the restricted geometry for the `nuclear-spin island', as well as the engineered uniform-coupling, may lead to a suppression of nuclear spin diffusion through dipolar coupling,\cite{2003_PRB_Sousa} prolonging nuclear-spin coherence times. Strategies based on a combination of isotopic engineering of the semiconductor substrate and electric manipulation of the electron wave-function should be a useful tool also beyond our specific setup, allowing for additional control of the electron and nuclear spin dynamics. 

Finally, we have focused here on quantum dots, which is partially motivated by recent experimental progress on electron shuttling.\cite{2017_npj_Quant_Info_Fujita,2018_Petta} The same ideas could be relevant to other platforms, e.g., donor impurities with high-spin nuclei,\cite{2010_PRL_George,2010_Nat_Mater_Morley,2018_PRE_Mourik,2019_arXiv_Asaad} where it would be important to assess the influence of quadrupolar interaction and strain.\cite{2015_PRL_Franke,2018_PRAppl_Pla,2018_PRL_Mansir}

We thank W. A. Coish, G. Burkard, and Wen Yang for helpful discussions. S. Chesi acknowledges support from the National Key Research and Development Program of China (Grant No. 2016YFA0301200), NSFC (Grants No. 11574025, No. 11750110428, and No. 1171101295) and NSAF (Grant No. U1930402). Y.-D. Wang acknowledges support from NSFC (Grant No. 11947302) and MOST (Grant No. 2017FA0304500).

\appendix

\section{Master equation of the quantum dot} \label{section_1a}

We present here the derivation of the master equation describing the stationary quatnum dot, i.e., based on Eq.~(\ref{H}) after tracing out the leads degrees of freedom. Restricting ourselves to the weak-tunneling regime,  we adopt the standard second-order Born-Markov approximation: \cite{2002_BOOK_Breuer,2012_BOOK_Blum}
\begin{equation}\label{master_eq_BM}
\dot{\tilde{\rho}}_{s}(t)=-\int_{0}^{\infty}d\tau\mathrm{Tr}_b\{[\tilde{H}_{T}(t),[\tilde{H}_{T}(t-\tau),\tilde{\rho}_{s}(t)\otimes\rho_{b}]]\},
\end{equation}
where $\rho_{s}(t)=\mathrm{Tr}_b\{\rho (t)\}$ is the reduced density matrix of the quantum dot, $\mathrm{Tr}_b\{...\}$ is the partial trace over the leads, and $\rho_b$ is the reduced density matrix of the leads with given chemical potentials $\mu_\alpha$ [see Eq.~(\ref{f_leads})]. The tilde indicates operators in the interaction picture, $\tilde{O}(t)=e^{i(H_0 +H_b)t}O(t)e^{-i(H_0+H_b) t}$. 

To evaluate Eq.~(\ref{master_eq_BM}) more explicitly, we use the exact eigenstates of $H_0$ given in Eq.~(\ref{eigenstates}). In this section, we indicate them as $|\xi\rangle$ (with energy $\epsilon_{\xi}$). In particular, we introduce the spectral decomposition $d^\dag_{\sigma}=\int_{-\infty}^{\infty}d\omega d^\dag_{\sigma}(\omega)$, where:\cite{2002_BOOK_Breuer} 
\begin{equation}
d_{\sigma}^{\dagger}(\omega)=[d_{\sigma}(\omega)]^{\dagger} = \sum_{\xi,\xi'}|\xi\rangle\langle\xi|d_{\sigma}^{\dagger}|\xi'\rangle\langle\xi'|\delta(\omega-\epsilon_{\xi}).\label{Sec II: d operator}
\end{equation}
It is then straightforward to write $H_T$ in the interaction picture and obtain:
\begin{align}\label{master_eq_nRWA}
\dot{\tilde\rho}_s(t) = \sum_{\sigma}\int d\omega  d\omega' \Big\{\Gamma_\mathrm{out}(\omega) 
\left[d_{\sigma}(\omega){\tilde\rho}_s(t), d_{\sigma}^{\dagger}(\omega')\right]  \nonumber \\
 +\Gamma_\mathrm{in}(\omega')\left[d_{\sigma}^{\dagger}(\omega'){\tilde\rho}_s(t),d_{\sigma}(\omega)\right]\Big\} e^{i(\omega'-\omega)t} +\mathrm{H.c.} ,
\end{align}
where we defined
\begin{align}\label{Gamma_omega} 
&\Gamma_\mathrm{out}(\omega)=\sum_{\alpha k}\int_0^{\infty} d\tau e^{i(\omega-\varepsilon_{\alpha k})\tau}|T_{\alpha k}|^{2}\left(1-f_{\alpha}(\varepsilon_{\alpha k})\right), \nonumber \\
&\Gamma_\mathrm{in}(\omega)=\sum_{\alpha k}\int_0^{\infty} d\tau e^{-i(\omega-\varepsilon_{\alpha k})\tau}|T_{\alpha k}|^{2}f_{\alpha}(\varepsilon_{\alpha k}).
\end{align}
Note that in Eq.~(\ref{Sec II: d operator}) the argument of the delta function contains $\epsilon_{\xi}$ instead of the transition frequency $\epsilon_{\xi}-\epsilon_{\xi'}$, simply because $|\xi'\rangle$ corresponds to an empty quantum dot and $\epsilon_{\xi'}=0$. After going back to the Shr\"odinger picture, the integrals over frequencies in Eq.~(\ref{master_eq_nRWA}) can be evaluated by introducing the operators $\Gamma_{\rm in/out}(H_0)$. It is easy to see that $\int d\omega \Gamma_{\rm out}(\omega)d_\sigma(\omega) = d_\sigma \Gamma_{\rm out}(H_0)  $, and similarly for other integrals of this type. Furthermore, introducing the Hermitian operators  $\gamma_{\rm in/out},\Delta_{\rm in/out}$:
\begin{align}\label{gamma_Delta_def}
 \Gamma_{\rm in/out}(H_0)  \equiv \frac{\gamma_{\rm in/out}}{2}+i \Delta_{\rm in/out},
\end{align}
and using that $\Delta_{\rm in/out} $ are approximately equal [$\Delta_{\rm in/out} \simeq \Delta_{\rm LS}$, see Eq.~(\ref{Delta_LS}) below] we arrive to:
\begin{align}\label{m_eq_1}
\dot{\rho}_s =  -i [H_0 ,   & \rho_s ]  +  \sum_\sigma   \left\{  \left[ d_\sigma \left(\frac{\gamma_{\rm out}}{2}+i\Delta_{\rm LS} \right)\rho_s, d_\sigma^\dag \right]\right. \nonumber \\
&\left. + \left[ \left(\frac{\gamma_{\rm in}}{2}+i\Delta_{\rm LS} \right) d^\dag_\sigma \rho_s, d_\sigma \right] + {\rm H.c.} \right\}.
\end{align}

We now give the explicit expressions of $\gamma_{\rm in/out}$ and $\Delta_{\rm LS}$, where as usual we transform $\sum_{k} \to \int d\varepsilon$ and compute the integrals assuming constant density of states and tunnel amplitudes. In this way, we obtain:
\begin{equation}\label{gamma_Delta}
\gamma_{\rm in}=\sum_\alpha \Gamma_\alpha \theta(\mu_\alpha - H_0), \quad  \gamma_{\rm out}=\sum_\alpha \Gamma_\alpha \theta(H_0-\mu_\alpha), 
\end{equation}
where the tunnel rates $\Gamma_{\alpha}$ are given in Eq.~(\ref{Gamma_lr}). For the Lamb-shift terms we have:
\begin{align}
&\Delta_{\rm in} = \sum_\alpha \Gamma_\alpha \bigg({\rm P} \int_{\mu_\alpha-\Lambda}^{\mu_\alpha}\frac{d\varepsilon}{2\pi}  \frac{1}{\varepsilon-H_0}\bigg), \nonumber \\ 
&\Delta_{\rm out} =\sum_\alpha \Gamma_\alpha \bigg({\rm P} \int_{\mu_\alpha}^{\mu_\alpha+\Lambda}   \frac{d\varepsilon}{2\pi} \frac{ 1}{H_0 - \varepsilon} \bigg),
\end{align}
where we supposed the $\alpha$ lead to have a bandwidth $2\Lambda$ around its chemical potential $\mu_\alpha$. In the limit of large $\Lambda$:
\begin{align}\label{Delta_LS}
\Delta_{\rm in/out} \simeq \Delta_{\rm LS} = \sum_\alpha \frac{\Gamma_\alpha}{2\pi} \ln \left( \frac{|\mu_\alpha - H_0|}{\Lambda} \right) .
\end{align}
Interestingly, the choice of the cutoff does not affect the evolution of $\rho_s$. In fact, by changing $\Lambda$, the right-hand side of  Eq.~(\ref{m_eq_1}) is modified by a term proportional to:
\begin{equation}
\sum_\sigma \left(  \left[ d_\sigma \rho_s, d_\sigma^\dag \right] +  \left[ d^\dag_\sigma \rho_s, d_\sigma \right] - {\rm H.c.}  \right) =
 \sum_\sigma  \left[ \rho_s ,\left\{d_\sigma^\dag , d_\sigma \right\}   \right],
\end{equation}
which is obviously zero since $\left\{d_\sigma^\dag , d_\sigma \right\} =1$.

So far, the main result of this section is Eq.~(\ref{m_eq_1}), which with uniform hyperfine coupling and fixed total angular momentum $I$ can be evaluated for a relatively large nuclear system. An example is given in Fig.~\ref{compare} of the main text. We emphasize that Eq.~(\ref{master_eq_BM}) and (\ref{m_eq_1}) are essentially equivalent, since the derivation of Eq.~(\ref{m_eq_1})  does not involve further approximations, except for standard assumptions on the leads density of states and tunnel amplitudes. Furthermore, we did not perform yet a rotating-wave approximation. For this reason, the dissipation of Eq.~(\ref{m_eq_1}) is not in the Lindblad form, and small unphysical effects can appear during the time evolution.

To obtain a master equation of the Lindblad type, we perform a partial rotating-wave approximation on Eq.~(\ref{m_eq_1}). We can also drop the Lamb shift, which usually has a small effect (see Fig.~\ref{compare}). To neglect fast-oscillating terms, we first express $\gamma_{\rm in/out}$ in terms of the projectors $P_\pm$ on the two well-separated bands of states. For example, for the bias configuration shown in the inset of Fig.~\ref{compare}:
\begin{equation}
\gamma_{\rm in} = (\Gamma_l + \Gamma_r)P_- + \Gamma_l P_+ , \quad
\gamma_{\rm out} =  \Gamma_r P_+ .
\end{equation}
Then, the projected fermionic operators $A_{\sigma\pm}$ [defined in Eq.~(\ref{Asigma_pm})] naturally appear in the master equation Eq.~(\ref{m_eq_1}). We can also use the fact that, since we always omit doubly occupied states, the $A_{\sigma \pm}$ provide a decomposition of the $d_\sigma$ operators: $d_\sigma = A_{\sigma +}+A_{\sigma -} $. Finally, based on the large energy separation between the $P_+$ and $P_-$ subspaces, we neglect in Eq.~(\ref{m_eq_1}) the cross-terms involving two bands simultaneously (i.e., the terms containing both  $A_{\sigma +}$ and $A_{\sigma -}$). This treatment lead to Eqs.~(\ref{ME_Lindblad_2b}) and (\ref{ME_Lindblad}) of the main text, where the dissipator is indeed of Lindblad type.

\section{Rate equations and small-$\eta$ expansion}\label{sec_rates}

An even simpler descrpition of the quantum dot dynamics is through rate equations. For our systems, the description through rate equations gives results which are in agreement with more sophisticated treatments. In some cases they are even equivalent to the evolution based on a full master equation. For example, a  detailed analysis of Eq.~(\ref{ME_Lindblad}) shows that for the initial state $|0\rangle \otimes|I,I \rangle$ the density matrix remains diagonal in the basis of the eigenstates. We will then consider the rate equations following Eq.~(\ref{ME_Lindblad}). By neglecting the coherence between all the eigenstates $|\xi \rangle$ of $H_0$, i.e., assuming $\langle \xi |\rho_s |\xi' \rangle \simeq P_\xi \delta_{\xi \xi'}$, we obtain:
\begin{align}\label{rate_eqs}
\dot{P}_{+,m} =&\Gamma _{l}\left( \alpha^2_{m}P_{0,m}+\beta_m^2P_{0,m+1}\right)-\Gamma _{r}P_{+,m} , \nonumber \\
\dot{P}_{-,m} =&\left( \Gamma _{r}+\Gamma _{l}\right)\left(\alpha_{m-1}^2 P_{0,m}+\beta_{m-1}^2 P_{0,m-1} \right),  \nonumber \\
\dot{P}_{0,m} =&\Gamma _{r}\left(\alpha_{m}^2 P_{+,m}+\beta_{m-1}^2 P_{+,m-1} \right)\nonumber \\
& -\left( 2\Gamma _{l}+\Gamma _{r}\beta_{m}^2+\Gamma
_{r}\alpha_{m-1}^2\right) P_{0,m}, 
\end{align}%
where $P_{\pm,m}=\langle \varphi_{I,m}^\pm |\rho_{s}|\varphi_{I,m}^\pm \rangle$ and $P_{0,m}=\langle 0,m|\rho_{s}|0,m\rangle$ are respectively the populations of the occupied and empty quantum dot. We recall here the notation $|0,m\rangle=|0\rangle\otimes|I,m\rangle$ and that $\alpha_{m}=\cos(\theta_{m}/2)$, $\beta_{m}=\sin(\theta_{m}/2)$, with the mixing angle given in Eq.~(\ref{theta_m}). 

The physical interpretation of Eq.~(\ref{rate_eqs}) is rather transparent, as the various contributions on the right-hand side can be associated to spin-conserving and spin-flipping tunnel events to/from the quantum dot: the terms proportional to $\beta_m^2$ correspond to tunneling events accompanied by a flip-flop process of the electron and nuclear spins. For such processes, the rates are suppressed by the square amplitude of the spin-flipped component in the quantum-dot eigenstates, see Eq.~(\ref{eigenstates}). Instead, the terms proportional to $\alpha_m^2$ are associated to processes when the nuclear spin flip does not take place. 

\begin{figure}
\begin{centering}
\includegraphics[width=0.45\textwidth]{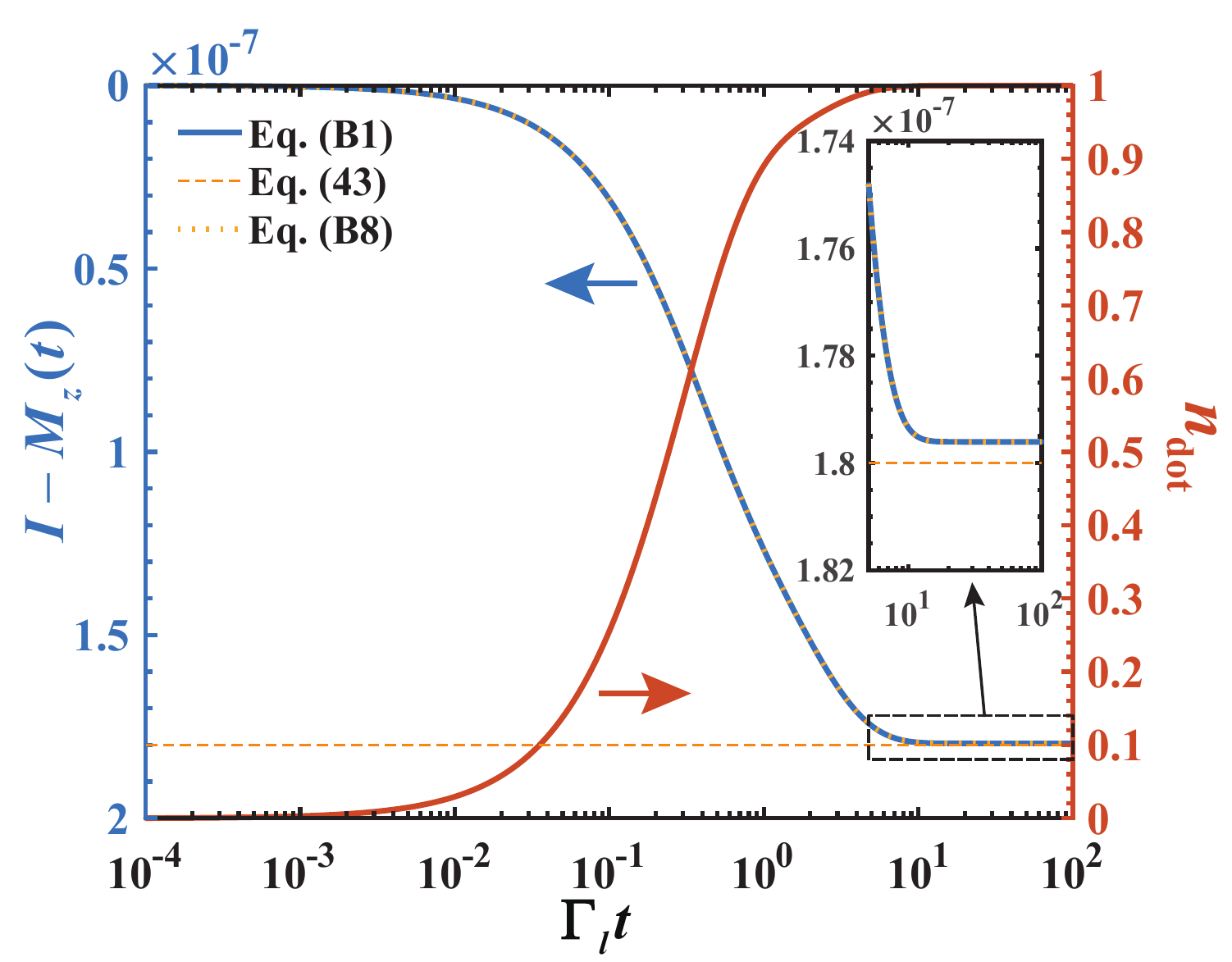}\label{rateq}
\par\end{centering}
\caption{Nuclear spins polarization dynamics from the rate equations Eq.~(\ref{rate_eqs}). The approximated results, i.e., dashed and dotted curves, agree well with the exact numerics (blue solid curve). The orange solid curve (right scale) is the corresponding quantum dot occupation, $n_{\rm dot}=\langle d_\uparrow^\dag d_\uparrow + d_\downarrow^\dag d_\downarrow \rangle$, which quickly saturates to $n_{\rm dot}=1$. Parameters used in the calculations (in unit of $\omega_0$): $N=12$, $I=N/2$, $\eta=10^{-4}$, and $\Gamma_l=\Gamma_r=0.1$.}
\end{figure}

To gain analytical insight into the rate equations (\ref{rate_eqs}) and obtain a simple analytical expression for the nuclear-spin magnetizaton $M_z(t)$, we take advatage of the small parameter $\eta$ and expand the populations perturbatively:
\begin{equation}
P_{s,m}=P_{s,m}^{(0)}+P_{s,m}^{(2)}+P_{s,m}^{(3)}+\ldots ,\label{P_expansion}
\end{equation}
where $P_{s,m}^{(k)}$ is proportional to $\eta^{k}$ (as we will see below, the $O(\eta)$ term is missing). The lowest-order result is obtained taking $\alpha_m^2\simeq 1$ and $\beta_m^2\simeq 0$ and gives the evolution in the absence of hyperfine interaction. For an initial state $\rho_s(0) = |0,I\rangle\langle 0, I |$, it is easy to obtain:
\begin{align}\label{P0}
& P_{+,I}^{(0)}(t)=\sqrt{\frac{\Gamma_{l}}{\Gamma_{l}+\Gamma_{r}}}e^{-(\Gamma_{l}+\Gamma_{r})t}\sinh\sqrt{\Gamma_{l}(\Gamma_{l}+\Gamma_{r})}t, \nonumber \\
& P_{-,I}^{(0)}(t)=1-e^{-(\Gamma_{l}+\Gamma_{r})t}\cosh\sqrt{\Gamma_{l}(\Gamma_{l}+\Gamma_{r})}t,
\end{align}
while $P_{0,I}^{(0)}=1-P_{+,I}^{(0)}-P_{-,I}^{(0)}$ and all other $P_{s,m}^{(0)}$ are zero. To obtain the higher-order contributions, we consider the expansion of $\beta^2_m$ (note that $\alpha_m^2=1-\beta^2_m$):
\begin{equation}\label{beta_eta}
\beta_{m}^{2}=g^{(2)}_{m}\eta^2+g^{(3)}_{m}\eta^{3}+\ldots,
\end{equation}
where $g^{(2)}_{m}=I(I+1)-m(m+1)$ and $g^{(3)}_{m}=-2(2m+1)g^{(2)}_{m}$. We see that the first correction is indeed of order $\eta^2$. More precisely, since $g^{(j)}_m \sim I^j$, the expansion parameter is $I\eta $. If we take $I\sim N$, the condition of validity becomes $A N/N_d\ll \omega_0$. By making use of Eqs.~(\ref{P0}) and (\ref{beta_eta}) in the rate equations, it is straightforward to obtain the equation of motions for $P_{s,m}^{(2)}$ and $P_{s,m}^{(3)}$. For example, defining $P_{m}^{(k)}=P_{0,m}^{(k)}+P_{+,m}^{(k)}+P_{-,m}^{(k)}$ we obtain the compact equation (with $j=2,3$):
\begin{equation}\label{P_eq}
\dot{P}_{I-1}^{(j)}(t)=-\dot{P}_{I}^{(j)}(t)=\eta^j g^{(j)}_{I-1}\Gamma_{l}P_{0,I}^{(0)}(t).
\end{equation}

It is also possible to apply the perturbative solution to the nuclear spin polarization:
\begin{equation}
M_{z}(t)=\sum_{m}m P_{0,m}+\sum_{\sigma,s,m,m'}m|\langle\sigma,m|\varphi_{I,m'}^{s}\rangle|^{2}P_{s,m'}.
\end{equation}
With the choice of initial state  $|0\rangle \otimes|I,I \rangle$, one immediately finds $M_{z}^{(0)}=I$. The $j=2,3$ corrections are:
\begin{equation}\label{M23}
M_{z}^{(j)}=\sum_{m}m P_{m}^{(j)}+\eta^{j} \left(g^{(j)}_I P_{+,I}^{(0)} -g^{(j)}_{I-1} P_{-,I}^{(0)}\right).
\end{equation}
As it turns out, in Eq.~(\ref{M23}) only $P_{I}^{(j)}$ and $P_{I-1}^{(j)}$ are different from zero, and using Eq.~(\ref{P_eq}) gives the nuclear-spin polarization:
\begin{align}\label{Mz_perturbative}
M_z(t)\simeq I & -2I\eta^2  \left( 1-2\eta(2I-1)\right) \nonumber \\ 
& \times \left(P_{-,I}^{(0)}(t)+\Gamma_l \int_0^t dt' P_{0,I}^{(0)}(t') \right),
\end{align}
which is plotted in Fig.~\ref{rateq} with and without the $O(\eta^3)$ contribution. We find that for small $\eta$ the lowest order correction is in excellent agreement with Eq.~(\ref{rate_eqs}). In the inset we show that including the third-order eliminates any visible discrepancy.

The stationary value can be obtained using $\int_0^\infty P_{0,I}^{(0)}(t) dt = (\Gamma_l + \Gamma_r)^{-1}$:
\begin{align}\label{Mz_infty}
M_{z} (t\to \infty) \simeq  
I -2I\eta^2 & \left( 1-2\eta(2I-1)\right) \frac{2\Gamma_{l}+\Gamma_r}{\Gamma_{l}+\Gamma_{r}},
\end{align}
which, omitting the $O(\eta^3)$ contribution, is Eq.~(\ref{Mz_approx}) of the main text.

\bibliographystyle{apsrev}
\bibliography{references}

\begin{thebibliography}{66}
\expandafter\ifx\csname natexlab\endcsname\relax\def\natexlab#1{#1}\fi
\expandafter\ifx\csname bibnamefont\endcsname\relax
  \def\bibnamefont#1{#1}\fi
\expandafter\ifx\csname bibfnamefont\endcsname\relax
  \def\bibfnamefont#1{#1}\fi
\expandafter\ifx\csname citenamefont\endcsname\relax
  \def\citenamefont#1{#1}\fi
\expandafter\ifx\csname url\endcsname\relax
  \def\url#1{\texttt{#1}}\fi
\expandafter\ifx\csname urlprefix\endcsname\relax\def\urlprefix{URL }\fi
\providecommand{\bibinfo}[2]{#2}
\providecommand{\eprint}[2][]{\url{#2}}

\bibitem[{\citenamefont{Coish and
  Baugh}(2009{\natexlab{a}})}]{2009_Phys_Stat_Sol_b_Coish}
\bibinfo{author}{\bibfnamefont{W.~A.} \bibnamefont{Coish}} \bibnamefont{and}
  \bibinfo{author}{\bibfnamefont{J.}~\bibnamefont{Baugh}},
  \bibinfo{journal}{Phys. Stat. Sol. B} \textbf{\bibinfo{volume}{246}},
  \bibinfo{pages}{2203} (\bibinfo{year}{2009}{\natexlab{a}}).

\bibitem[{\citenamefont{Yang et~al.}(2017)\citenamefont{Yang, Ma, and
  Liu}}]{2017_Rep_Phys_Yang}
\bibinfo{author}{\bibfnamefont{W.}~\bibnamefont{Yang}},
  \bibinfo{author}{\bibfnamefont{W.-L.} \bibnamefont{Ma}}, \bibnamefont{and}
  \bibinfo{author}{\bibfnamefont{R.-B.} \bibnamefont{Liu}},
  \bibinfo{journal}{Rep. Prog. Phys.} \textbf{\bibinfo{volume}{80}},
  \bibinfo{pages}{016001} (\bibinfo{year}{2017}).

\bibitem[{\citenamefont{Koppens et~al.}(2007)\citenamefont{Koppens, Klauser,
  Coish, Nowack, Kouwenhoven, Loss, and Vandersypen}}]{2007_PRL_Koppens}
\bibinfo{author}{\bibfnamefont{F.~H.~L.} \bibnamefont{Koppens}},
  \bibinfo{author}{\bibfnamefont{D.}~\bibnamefont{Klauser}},
  \bibinfo{author}{\bibfnamefont{W.~A.} \bibnamefont{Coish}},
  \bibinfo{author}{\bibfnamefont{K.~C.} \bibnamefont{Nowack}},
  \bibinfo{author}{\bibfnamefont{L.~P.} \bibnamefont{Kouwenhoven}},
  \bibinfo{author}{\bibfnamefont{D.}~\bibnamefont{Loss}}, \bibnamefont{and}
  \bibinfo{author}{\bibfnamefont{L.~M.~K.} \bibnamefont{Vandersypen}},
  \bibinfo{journal}{Phys. Rev. Lett.} \textbf{\bibinfo{volume}{99}},
  \bibinfo{pages}{106803} (\bibinfo{year}{2007}).

\bibitem[{\citenamefont{Chesi et~al.}(2016)\citenamefont{Chesi, Yang, and
  Loss}}]{2016_PRL_Chesi}
\bibinfo{author}{\bibfnamefont{S.}~\bibnamefont{Chesi}},
  \bibinfo{author}{\bibfnamefont{L.-P.} \bibnamefont{Yang}}, \bibnamefont{and}
  \bibinfo{author}{\bibfnamefont{D.}~\bibnamefont{Loss}},
  \bibinfo{journal}{Phys. Rev. Lett.} \textbf{\bibinfo{volume}{116}},
  \bibinfo{pages}{066806} (\bibinfo{year}{2016}).

\bibitem[{\citenamefont{Bluhm et~al.}(2011)\citenamefont{Bluhm, Foletti, Neder,
  Rudner, Mahalu, Umansky, and Yacoby}}]{2011_Nat_Phys_Bluhm}
\bibinfo{author}{\bibfnamefont{H.}~\bibnamefont{Bluhm}},
  \bibinfo{author}{\bibfnamefont{S.}~\bibnamefont{Foletti}},
  \bibinfo{author}{\bibfnamefont{I.}~\bibnamefont{Neder}},
  \bibinfo{author}{\bibfnamefont{M.}~\bibnamefont{Rudner}},
  \bibinfo{author}{\bibfnamefont{D.}~\bibnamefont{Mahalu}},
  \bibinfo{author}{\bibfnamefont{V.}~\bibnamefont{Umansky}}, \bibnamefont{and}
  \bibinfo{author}{\bibfnamefont{A.}~\bibnamefont{Yacoby}},
  \bibinfo{journal}{Nat. Phys.} \textbf{\bibinfo{volume}{7}},
  \bibinfo{pages}{109} (\bibinfo{year}{2011}).

\bibitem[{\citenamefont{Malinowski
  et~al.}(2017{\natexlab{a}})\citenamefont{Malinowski, Martins, Nissen, Barnes,
  Cywi{\'n}ski, Rudner, Fallahi, Gardner, Manfra, Marcus
  et~al.}}]{2016_Nat_Nano_Malinowski}
\bibinfo{author}{\bibfnamefont{F.~K.} \bibnamefont{Malinowski}},
  \bibinfo{author}{\bibfnamefont{F.}~\bibnamefont{Martins}},
  \bibinfo{author}{\bibfnamefont{P.~D.} \bibnamefont{Nissen}},
  \bibinfo{author}{\bibfnamefont{E.}~\bibnamefont{Barnes}},
  \bibinfo{author}{\bibfnamefont{{\L}.}~\bibnamefont{Cywi{\'n}ski}},
  \bibinfo{author}{\bibfnamefont{M.~S.} \bibnamefont{Rudner}},
  \bibinfo{author}{\bibfnamefont{S.}~\bibnamefont{Fallahi}},
  \bibinfo{author}{\bibfnamefont{G.~C.} \bibnamefont{Gardner}},
  \bibinfo{author}{\bibfnamefont{M.~J.} \bibnamefont{Manfra}},
  \bibinfo{author}{\bibfnamefont{C.~M.} \bibnamefont{Marcus}},
  \bibnamefont{et~al.}, \bibinfo{journal}{Nat. Nano.}
  \textbf{\bibinfo{volume}{12}}, \bibinfo{pages}{16}
  (\bibinfo{year}{2017}{\natexlab{a}}).

\bibitem[{\citenamefont{Bluhm et~al.}(2010)\citenamefont{Bluhm, Foletti,
  Mahalu, Umansky, and Yacoby}}]{2010_PRL_Bluhm}
\bibinfo{author}{\bibfnamefont{H.}~\bibnamefont{Bluhm}},
  \bibinfo{author}{\bibfnamefont{S.}~\bibnamefont{Foletti}},
  \bibinfo{author}{\bibfnamefont{D.}~\bibnamefont{Mahalu}},
  \bibinfo{author}{\bibfnamefont{V.}~\bibnamefont{Umansky}}, \bibnamefont{and}
  \bibinfo{author}{\bibfnamefont{A.}~\bibnamefont{Yacoby}},
  \bibinfo{journal}{Phys. Rev. Lett.} \textbf{\bibinfo{volume}{105}},
  \bibinfo{pages}{216803} (\bibinfo{year}{2010}).

\bibitem[{\citenamefont{Shulman et~al.}(2014)\citenamefont{Shulman, Harvey,
  Nichol, Bartlett, Doherty, Umansky, and Yacoby}}]{2014_Nat_Commun_Shulman}
\bibinfo{author}{\bibfnamefont{M.~D.} \bibnamefont{Shulman}},
  \bibinfo{author}{\bibfnamefont{S.~P.} \bibnamefont{Harvey}},
  \bibinfo{author}{\bibfnamefont{J.~M.} \bibnamefont{Nichol}},
  \bibinfo{author}{\bibfnamefont{S.~D.} \bibnamefont{Bartlett}},
  \bibinfo{author}{\bibfnamefont{A.~C.} \bibnamefont{Doherty}},
  \bibinfo{author}{\bibfnamefont{V.}~\bibnamefont{Umansky}}, \bibnamefont{and}
  \bibinfo{author}{\bibfnamefont{A.}~\bibnamefont{Yacoby}},
  \bibinfo{journal}{Nat. Commun.} \textbf{\bibinfo{volume}{5}},
  \bibinfo{pages}{5156} (\bibinfo{year}{2014}).

\bibitem[{\citenamefont{Delbecq et~al.}(2016)\citenamefont{Delbecq, Nakajima,
  Stano, Otsuka, Amaha, Yoneda, Takeda, Allison, Ludwig, Wieck
  et~al.}}]{2016_PRL_Delbecq}
\bibinfo{author}{\bibfnamefont{M.~R.} \bibnamefont{Delbecq}},
  \bibinfo{author}{\bibfnamefont{T.}~\bibnamefont{Nakajima}},
  \bibinfo{author}{\bibfnamefont{P.}~\bibnamefont{Stano}},
  \bibinfo{author}{\bibfnamefont{T.}~\bibnamefont{Otsuka}},
  \bibinfo{author}{\bibfnamefont{S.}~\bibnamefont{Amaha}},
  \bibinfo{author}{\bibfnamefont{J.}~\bibnamefont{Yoneda}},
  \bibinfo{author}{\bibfnamefont{K.}~\bibnamefont{Takeda}},
  \bibinfo{author}{\bibfnamefont{G.}~\bibnamefont{Allison}},
  \bibinfo{author}{\bibfnamefont{A.}~\bibnamefont{Ludwig}},
  \bibinfo{author}{\bibfnamefont{A.~D.} \bibnamefont{Wieck}},
  \bibnamefont{et~al.}, \bibinfo{journal}{Phys. Rev. Lett.}
  \textbf{\bibinfo{volume}{116}}, \bibinfo{pages}{046802}
  (\bibinfo{year}{2016}).

\bibitem[{\citenamefont{Malinowski
  et~al.}(2017{\natexlab{b}})\citenamefont{Malinowski, Martins,
  Cywi\ifmmode~\acute{n}\else \'{n}\fi{}ski, Rudner, Nissen, Fallahi, Gardner,
  Manfra, Marcus, and Kuemmeth}}]{2017_PRL_Malinowski}
\bibinfo{author}{\bibfnamefont{F.~K.} \bibnamefont{Malinowski}},
  \bibinfo{author}{\bibfnamefont{F.}~\bibnamefont{Martins}},
  \bibinfo{author}{\bibfnamefont{L.}~\bibnamefont{Cywi\ifmmode~\acute{n}\else
  \'{n}\fi{}ski}}, \bibinfo{author}{\bibfnamefont{M.~S.} \bibnamefont{Rudner}},
  \bibinfo{author}{\bibfnamefont{P.~D.} \bibnamefont{Nissen}},
  \bibinfo{author}{\bibfnamefont{S.}~\bibnamefont{Fallahi}},
  \bibinfo{author}{\bibfnamefont{G.~C.} \bibnamefont{Gardner}},
  \bibinfo{author}{\bibfnamefont{M.~J.} \bibnamefont{Manfra}},
  \bibinfo{author}{\bibfnamefont{C.~M.} \bibnamefont{Marcus}},
  \bibnamefont{and} \bibinfo{author}{\bibfnamefont{F.}~\bibnamefont{Kuemmeth}},
  \bibinfo{journal}{Phys. Rev. Lett.} \textbf{\bibinfo{volume}{118}},
  \bibinfo{pages}{177702} (\bibinfo{year}{2017}{\natexlab{b}}).

\bibitem[{\citenamefont{Pla et~al.}(2013)\citenamefont{Pla, Tan, Dehollain,
  Lim, Morton, Zwanenburg, Jamieson, Dzurak, and Morello}}]{2013_Nature_Pla}
\bibinfo{author}{\bibfnamefont{J.~J.} \bibnamefont{Pla}},
  \bibinfo{author}{\bibfnamefont{K.~Y.} \bibnamefont{Tan}},
  \bibinfo{author}{\bibfnamefont{J.~P.} \bibnamefont{Dehollain}},
  \bibinfo{author}{\bibfnamefont{W.~H.} \bibnamefont{Lim}},
  \bibinfo{author}{\bibfnamefont{J.~J.~L.} \bibnamefont{Morton}},
  \bibinfo{author}{\bibfnamefont{F.~A.} \bibnamefont{Zwanenburg}},
  \bibinfo{author}{\bibfnamefont{D.~N.} \bibnamefont{Jamieson}},
  \bibinfo{author}{\bibfnamefont{A.~S.} \bibnamefont{Dzurak}},
  \bibnamefont{and} \bibinfo{author}{\bibfnamefont{A.}~\bibnamefont{Morello}},
  \bibinfo{journal}{Nature} \textbf{\bibinfo{volume}{496}},
  \bibinfo{pages}{334} (\bibinfo{year}{2013}).

\bibitem[{\citenamefont{van~der Sar et~al.}(2012)\citenamefont{van~der Sar,
  Wang, Blok, Bernien, Taminiau, Toyli, Lidar, Awschalom, Hanson, and
  Dobrovitski}}]{2012_Nature_van_der_Sar}
\bibinfo{author}{\bibfnamefont{T.}~\bibnamefont{van~der Sar}},
  \bibinfo{author}{\bibfnamefont{Z.~H.} \bibnamefont{Wang}},
  \bibinfo{author}{\bibfnamefont{M.~S.} \bibnamefont{Blok}},
  \bibinfo{author}{\bibfnamefont{H.}~\bibnamefont{Bernien}},
  \bibinfo{author}{\bibfnamefont{T.~H.} \bibnamefont{Taminiau}},
  \bibinfo{author}{\bibfnamefont{D.~M.} \bibnamefont{Toyli}},
  \bibinfo{author}{\bibfnamefont{D.~A.} \bibnamefont{Lidar}},
  \bibinfo{author}{\bibfnamefont{D.~D.} \bibnamefont{Awschalom}},
  \bibinfo{author}{\bibfnamefont{R.}~\bibnamefont{Hanson}}, \bibnamefont{and}
  \bibinfo{author}{\bibfnamefont{V.~V.} \bibnamefont{Dobrovitski}},
  \bibinfo{journal}{Nature} \textbf{\bibinfo{volume}{484}}, \bibinfo{pages}{82}
  (\bibinfo{year}{2012}).

\bibitem[{\citenamefont{Dicke}(1954)}]{1954_PR_Dicke}
\bibinfo{author}{\bibfnamefont{R.~H.} \bibnamefont{Dicke}},
  \bibinfo{journal}{Phys. Rev.} \textbf{\bibinfo{volume}{93}},
  \bibinfo{pages}{99} (\bibinfo{year}{1954}).

\bibitem[{\citenamefont{Degiorgio and Ghielmetti}(1971)}]{1971_PRA_Degiorgio}
\bibinfo{author}{\bibfnamefont{V.}~\bibnamefont{Degiorgio}} \bibnamefont{and}
  \bibinfo{author}{\bibfnamefont{F.}~\bibnamefont{Ghielmetti}},
  \bibinfo{journal}{Phys. Rev. A} \textbf{\bibinfo{volume}{4}},
  \bibinfo{pages}{2415} (\bibinfo{year}{1971}).

\bibitem[{\citenamefont{Gross and Haroche}(1982)}]{1982_Phys_Rep_Gross}
\bibinfo{author}{\bibfnamefont{M.}~\bibnamefont{Gross}} \bibnamefont{and}
  \bibinfo{author}{\bibfnamefont{S.}~\bibnamefont{Haroche}},
  \bibinfo{journal}{Phys. Rep.} \textbf{\bibinfo{volume}{93}},
  \bibinfo{pages}{301} (\bibinfo{year}{1982}).

\bibitem[{\citenamefont{Kessler et~al.}(2010)\citenamefont{Kessler, Yelin,
  Lukin, Cirac, and Giedke}}]{2010_PRL_Kessler}
\bibinfo{author}{\bibfnamefont{E.~M.} \bibnamefont{Kessler}},
  \bibinfo{author}{\bibfnamefont{S.}~\bibnamefont{Yelin}},
  \bibinfo{author}{\bibfnamefont{M.~D.} \bibnamefont{Lukin}},
  \bibinfo{author}{\bibfnamefont{J.~I.} \bibnamefont{Cirac}}, \bibnamefont{and}
  \bibinfo{author}{\bibfnamefont{G.}~\bibnamefont{Giedke}},
  \bibinfo{journal}{Phys. Rev. Lett.} \textbf{\bibinfo{volume}{104}},
  \bibinfo{pages}{143601} (\bibinfo{year}{2010}).

\bibitem[{\citenamefont{He et~al.}(2019)\citenamefont{He, Chesi, Lin, and
  Guan}}]{2019_PRB_He}
\bibinfo{author}{\bibfnamefont{W.-B.} \bibnamefont{He}},
  \bibinfo{author}{\bibfnamefont{S.}~\bibnamefont{Chesi}},
  \bibinfo{author}{\bibfnamefont{H.-Q.} \bibnamefont{Lin}}, \bibnamefont{and}
  \bibinfo{author}{\bibfnamefont{X.-W.} \bibnamefont{Guan}},
  \bibinfo{journal}{Phys. Rev. B} \textbf{\bibinfo{volume}{99}},
  \bibinfo{pages}{174308} (\bibinfo{year}{2019}).

\bibitem[{\citenamefont{Eto et~al.}(2004)\citenamefont{Eto, Ashiwa, and
  Murata}}]{2004_JPSJ_Eto}
\bibinfo{author}{\bibfnamefont{M.}~\bibnamefont{Eto}},
  \bibinfo{author}{\bibfnamefont{T.}~\bibnamefont{Ashiwa}}, \bibnamefont{and}
  \bibinfo{author}{\bibfnamefont{M.}~\bibnamefont{Murata}},
  \bibinfo{journal}{J. Phys. Soc. Jpn.} \textbf{\bibinfo{volume}{73}},
  \bibinfo{pages}{307} (\bibinfo{year}{2004}).

\bibitem[{\citenamefont{Schuetz et~al.}(2012)\citenamefont{Schuetz, Kessler,
  Cirac, and Giedke}}]{2012_PRB_Schuetz}
\bibinfo{author}{\bibfnamefont{M.~J.~A.} \bibnamefont{Schuetz}},
  \bibinfo{author}{\bibfnamefont{E.~M.} \bibnamefont{Kessler}},
  \bibinfo{author}{\bibfnamefont{J.~I.} \bibnamefont{Cirac}}, \bibnamefont{and}
  \bibinfo{author}{\bibfnamefont{G.}~\bibnamefont{Giedke}},
  \bibinfo{journal}{Phys. Rev. B} \textbf{\bibinfo{volume}{86}},
  \bibinfo{pages}{085322} (\bibinfo{year}{2012}).

\bibitem[{\citenamefont{Chesi and Coish}(2015)}]{2015_PRB_Stefano}
\bibinfo{author}{\bibfnamefont{S.}~\bibnamefont{Chesi}} \bibnamefont{and}
  \bibinfo{author}{\bibfnamefont{W.~A.} \bibnamefont{Coish}},
  \bibinfo{journal}{Phys. Rev. B} \textbf{\bibinfo{volume}{91}},
  \bibinfo{pages}{245306} (\bibinfo{year}{2015}).

\bibitem[{\citenamefont{Gorelik
  et~al.}(1998{\natexlab{a}})\citenamefont{Gorelik, Isacsson, Voinova, Kasemo,
  Shekhter, and Jonson}}]{1998_PRL_Gorelik}
\bibinfo{author}{\bibfnamefont{L.~Y.} \bibnamefont{Gorelik}},
  \bibinfo{author}{\bibfnamefont{A.}~\bibnamefont{Isacsson}},
  \bibinfo{author}{\bibfnamefont{M.~V.} \bibnamefont{Voinova}},
  \bibinfo{author}{\bibfnamefont{B.}~\bibnamefont{Kasemo}},
  \bibinfo{author}{\bibfnamefont{R.~I.} \bibnamefont{Shekhter}},
  \bibnamefont{and} \bibinfo{author}{\bibfnamefont{M.}~\bibnamefont{Jonson}},
  \bibinfo{journal}{Phys. Rev. Lett.} \textbf{\bibinfo{volume}{80}},
  \bibinfo{pages}{4526} (\bibinfo{year}{1998}{\natexlab{a}}).

\bibitem[{\citenamefont{Gorelik
  et~al.}(1998{\natexlab{b}})\citenamefont{Gorelik, Isacsson, Voinova, Kasemo,
  and Jonson}}]{1998_Physica_B_Gorelik}
\bibinfo{author}{\bibfnamefont{L.}~\bibnamefont{Gorelik}},
  \bibinfo{author}{\bibfnamefont{A.}~\bibnamefont{Isacsson}},
  \bibinfo{author}{\bibfnamefont{M.}~\bibnamefont{Voinova}},
  \bibinfo{author}{\bibfnamefont{R.}~\bibnamefont{Kasemo},
  \bibfnamefont{B.~Shekhter}}, \bibnamefont{and}
  \bibinfo{author}{\bibfnamefont{M.}~\bibnamefont{Jonson}},
  \bibinfo{journal}{Physica B} \textbf{\bibinfo{volume}{80}},
  \bibinfo{pages}{4526} (\bibinfo{year}{1998}{\natexlab{b}}).

\bibitem[{\citenamefont{Isacsson et~al.}(1998)\citenamefont{Isacsson, Gorelik,
  Voinova, Kasemo, Shekhter, and Jonson}}]{1998_Physica_B_Isacsson}
\bibinfo{author}{\bibfnamefont{A.}~\bibnamefont{Isacsson}},
  \bibinfo{author}{\bibfnamefont{L.~Y.} \bibnamefont{Gorelik}},
  \bibinfo{author}{\bibfnamefont{M.~V.} \bibnamefont{Voinova}},
  \bibinfo{author}{\bibfnamefont{B.}~\bibnamefont{Kasemo}},
  \bibinfo{author}{\bibfnamefont{R.~I.} \bibnamefont{Shekhter}},
  \bibnamefont{and} \bibinfo{author}{\bibfnamefont{M.}~\bibnamefont{Jonson}},
  \bibinfo{journal}{Physica B} \textbf{\bibinfo{volume}{64}},
  \bibinfo{pages}{035326} (\bibinfo{year}{1998}).

\bibitem[{\citenamefont{Gorelik
  et~al.}(1998{\natexlab{c}})\citenamefont{Gorelik, Kulinich, Galperin,
  Shekhter, and Jonson}}]{1998_PU_Gorelik}
\bibinfo{author}{\bibfnamefont{L.~.} \bibnamefont{Gorelik}},
  \bibinfo{author}{\bibfnamefont{S.}~\bibnamefont{Kulinich}},
  \bibinfo{author}{\bibfnamefont{Y.}~\bibnamefont{Galperin}},
  \bibinfo{author}{\bibfnamefont{R.~I.} \bibnamefont{Shekhter}},
  \bibnamefont{and} \bibinfo{author}{\bibfnamefont{M.}~\bibnamefont{Jonson}},
  \bibinfo{journal}{Phys.-Usp.} \textbf{\bibinfo{volume}{41}},
  \bibinfo{pages}{178} (\bibinfo{year}{1998}{\natexlab{c}}).

\bibitem[{\citenamefont{Gorelik et~al.}(2001)\citenamefont{Gorelik, Isacsson,
  Galperin, Shekhter, and Jonson}}]{2001_Nature_Gorelik}
\bibinfo{author}{\bibfnamefont{L.~Y.} \bibnamefont{Gorelik}},
  \bibinfo{author}{\bibfnamefont{A.}~\bibnamefont{Isacsson}},
  \bibinfo{author}{\bibfnamefont{Y.~M.} \bibnamefont{Galperin}},
  \bibinfo{author}{\bibfnamefont{R.~I.} \bibnamefont{Shekhter}},
  \bibnamefont{and} \bibinfo{author}{\bibfnamefont{M.}~\bibnamefont{Jonson}},
  \bibinfo{journal}{Nature} \textbf{\bibinfo{volume}{411}},
  \bibinfo{pages}{454} (\bibinfo{year}{2001}).

\bibitem[{\citenamefont{Fujita et~al.}(2017)\citenamefont{Fujita, Baart,
  Reichl, Wegscheider, and Vandersypen}}]{2017_npj_Quant_Info_Fujita}
\bibinfo{author}{\bibfnamefont{T.}~\bibnamefont{Fujita}},
  \bibinfo{author}{\bibfnamefont{T.~A.} \bibnamefont{Baart}},
  \bibinfo{author}{\bibfnamefont{C.}~\bibnamefont{Reichl}},
  \bibinfo{author}{\bibfnamefont{W.}~\bibnamefont{Wegscheider}},
  \bibnamefont{and} \bibinfo{author}{\bibfnamefont{L.~M.~K.}
  \bibnamefont{Vandersypen}}, \bibinfo{journal}{npj Quantum Inf.}
  \textbf{\bibinfo{volume}{3}}, \bibinfo{pages}{22} (\bibinfo{year}{2017}).

\bibitem[{\citenamefont{Mills et~al.}(2019)\citenamefont{Mills, Zajac, Gullans,
  Schupp, Hazard, and Petta}}]{2018_Petta}
\bibinfo{author}{\bibfnamefont{A.~R.} \bibnamefont{Mills}},
  \bibinfo{author}{\bibfnamefont{D.~M.} \bibnamefont{Zajac}},
  \bibinfo{author}{\bibfnamefont{M.~J.} \bibnamefont{Gullans}},
  \bibinfo{author}{\bibfnamefont{F.~J.} \bibnamefont{Schupp}},
  \bibinfo{author}{\bibfnamefont{T.~M.} \bibnamefont{Hazard}},
  \bibnamefont{and} \bibinfo{author}{\bibfnamefont{J.~R.} \bibnamefont{Petta}},
  \bibinfo{journal}{Nat. Commun.} \textbf{\bibinfo{volume}{10}},
  \bibinfo{pages}{1063} (\bibinfo{year}{2019}).

\bibitem[{\citenamefont{Park et~al.}(2000)\citenamefont{Park, Park, Lim,
  Anderson, Alivisatos, and McEuen}}]{2000_Nature_Park}
\bibinfo{author}{\bibfnamefont{H.~K.} \bibnamefont{Park}},
  \bibinfo{author}{\bibfnamefont{J.}~\bibnamefont{Park}},
  \bibinfo{author}{\bibfnamefont{A.~K.~L.} \bibnamefont{Lim}},
  \bibinfo{author}{\bibfnamefont{E.~H.} \bibnamefont{Anderson}},
  \bibinfo{author}{\bibfnamefont{A.~P.} \bibnamefont{Alivisatos}},
  \bibnamefont{and} \bibinfo{author}{\bibfnamefont{P.~L.}
  \bibnamefont{McEuen}}, \bibinfo{journal}{Nature}
  \textbf{\bibinfo{volume}{407}}, \bibinfo{pages}{57} (\bibinfo{year}{2000}).

\bibitem[{\citenamefont{Scheible and Blick}(2004)}]{2004_APL_Scheible}
\bibinfo{author}{\bibfnamefont{D.~V.} \bibnamefont{Scheible}} \bibnamefont{and}
  \bibinfo{author}{\bibfnamefont{R.~H.} \bibnamefont{Blick}},
  \bibinfo{journal}{Appl. Phys. Lett.} \textbf{\bibinfo{volume}{84}},
  \bibinfo{pages}{4632} (\bibinfo{year}{2004}).

\bibitem[{\citenamefont{Novotny et~al.}(2004)\citenamefont{Novotny, Donarini,
  Flindt, and Jauho}}]{2004_PRL_Novotny}
\bibinfo{author}{\bibfnamefont{T.}~\bibnamefont{Novotny}},
  \bibinfo{author}{\bibfnamefont{A.}~\bibnamefont{Donarini}},
  \bibinfo{author}{\bibfnamefont{C.}~\bibnamefont{Flindt}}, \bibnamefont{and}
  \bibinfo{author}{\bibfnamefont{A.~P.} \bibnamefont{Jauho}},
  \bibinfo{journal}{Phys. Rev. Lett.} \textbf{\bibinfo{volume}{92}},
  \bibinfo{pages}{248302} (\bibinfo{year}{2004}).

\bibitem[{\citenamefont{Pistolesi and Fazio}(2005)}]{2005_PRL_Pistolesi}
\bibinfo{author}{\bibfnamefont{F.}~\bibnamefont{Pistolesi}} \bibnamefont{and}
  \bibinfo{author}{\bibfnamefont{R.}~\bibnamefont{Fazio}},
  \bibinfo{journal}{Phys. Rev. Lett.} \textbf{\bibinfo{volume}{94}},
  \bibinfo{pages}{036806} (\bibinfo{year}{2005}).

\bibitem[{\citenamefont{Donarini et~al.}(2005)\citenamefont{Donarini, Novotny,
  and Jauho}}]{2005_NJP_Donarini}
\bibinfo{author}{\bibfnamefont{A.}~\bibnamefont{Donarini}},
  \bibinfo{author}{\bibfnamefont{T.}~\bibnamefont{Novotny}}, \bibnamefont{and}
  \bibinfo{author}{\bibfnamefont{A.~P.} \bibnamefont{Jauho}},
  \bibinfo{journal}{New J. Phys.} \textbf{\bibinfo{volume}{7}},
  \bibinfo{pages}{237} (\bibinfo{year}{2005}).

\bibitem[{\citenamefont{Pistolesi}(2004)}]{2004_PRB_Pistolesi}
\bibinfo{author}{\bibfnamefont{F.}~\bibnamefont{Pistolesi}},
  \bibinfo{journal}{Phys. Rev. B} \textbf{\bibinfo{volume}{69}},
  \bibinfo{pages}{245409} (\bibinfo{year}{2004}).

\bibitem[{\citenamefont{Romito and Nazarov}(2004)}]{2004_PRB_Romito}
\bibinfo{author}{\bibfnamefont{A.}~\bibnamefont{Romito}} \bibnamefont{and}
  \bibinfo{author}{\bibfnamefont{Y.~V.} \bibnamefont{Nazarov}},
  \bibinfo{journal}{Phys. Rev. B} \textbf{\bibinfo{volume}{70}},
  \bibinfo{pages}{212509} (\bibinfo{year}{2004}).

\bibitem[{\citenamefont{Khaetskii et~al.}(2003)\citenamefont{Khaetskii, Loss,
  and Glazman}}]{2003_PRB_Khaetskii}
\bibinfo{author}{\bibfnamefont{A.}~\bibnamefont{Khaetskii}},
  \bibinfo{author}{\bibfnamefont{D.}~\bibnamefont{Loss}}, \bibnamefont{and}
  \bibinfo{author}{\bibfnamefont{L.}~\bibnamefont{Glazman}},
  \bibinfo{journal}{Phys. Rev. B} \textbf{\bibinfo{volume}{67}},
  \bibinfo{pages}{195329} (\bibinfo{year}{2003}).

\bibitem[{\citenamefont{Coish et~al.}(2007)\citenamefont{Coish, Loss,
  Yuzbashyan, and Altshuler}}]{2007_JAP_Coish}
\bibinfo{author}{\bibfnamefont{W.~A.} \bibnamefont{Coish}},
  \bibinfo{author}{\bibfnamefont{D.}~\bibnamefont{Loss}},
  \bibinfo{author}{\bibfnamefont{E.~A.} \bibnamefont{Yuzbashyan}},
  \bibnamefont{and} \bibinfo{author}{\bibfnamefont{B.~L.}
  \bibnamefont{Altshuler}}, \bibinfo{journal}{J. Appl. Phys.}
  \textbf{\bibinfo{volume}{101}}, \bibinfo{pages}{081715}
  (\bibinfo{year}{2007}).

\bibitem[{\citenamefont{Zhang et~al.}(2006)\citenamefont{Zhang, Dobrovitski,
  Al-Hassanieh, Dagotto, and Harmon}}]{2006_PRB_Zhang}
\bibinfo{author}{\bibfnamefont{W.}~\bibnamefont{Zhang}},
  \bibinfo{author}{\bibfnamefont{V.~V.} \bibnamefont{Dobrovitski}},
  \bibinfo{author}{\bibfnamefont{K.~A.} \bibnamefont{Al-Hassanieh}},
  \bibinfo{author}{\bibfnamefont{E.}~\bibnamefont{Dagotto}}, \bibnamefont{and}
  \bibinfo{author}{\bibfnamefont{B.~N.} \bibnamefont{Harmon}},
  \bibinfo{journal}{Phys. Rev. B} \textbf{\bibinfo{volume}{74}},
  \bibinfo{pages}{205313} (\bibinfo{year}{2006}).

\bibitem[{\citenamefont{Chirolli and Burkard}(2008)}]{2008_Adv_Phys_Chirolli}
\bibinfo{author}{\bibfnamefont{L.}~\bibnamefont{Chirolli}} \bibnamefont{and}
  \bibinfo{author}{\bibfnamefont{G.}~\bibnamefont{Burkard}},
  \bibinfo{journal}{Adv. Phys.} \textbf{\bibinfo{volume}{57}},
  \bibinfo{pages}{225} (\bibinfo{year}{2008}).

\bibitem[{\citenamefont{Coish and Baugh}(2009{\natexlab{b}})}]{2009_PSSB_Coish}
\bibinfo{author}{\bibfnamefont{W.~A.} \bibnamefont{Coish}} \bibnamefont{and}
  \bibinfo{author}{\bibfnamefont{J.}~\bibnamefont{Baugh}},
  \bibinfo{journal}{Phys. Status Solidi B} \textbf{\bibinfo{volume}{246}},
  \bibinfo{pages}{2203} (\bibinfo{year}{2009}{\natexlab{b}}).

\bibitem[{\citenamefont{Lauhon et~al.}(2002)\citenamefont{Lauhon, Gudiksen,
  Wang, and Lieber}}]{2002_Nature_Lauhon}
\bibinfo{author}{\bibfnamefont{L.~J.} \bibnamefont{Lauhon}},
  \bibinfo{author}{\bibfnamefont{M.~S.} \bibnamefont{Gudiksen}},
  \bibinfo{author}{\bibfnamefont{D.~L.} \bibnamefont{Wang}}, \bibnamefont{and}
  \bibinfo{author}{\bibfnamefont{C.~M.} \bibnamefont{Lieber}},
  \bibinfo{journal}{Nature} \textbf{\bibinfo{volume}{420}}, \bibinfo{pages}{57}
  (\bibinfo{year}{2002}).

\bibitem[{\citenamefont{Moutanabbir et~al.}(2011)\citenamefont{Moutanabbir,
  Isheim, Seidman, Kawamura, and Itoh}}]{2011_APL_Moutanabbir}
\bibinfo{author}{\bibfnamefont{O.}~\bibnamefont{Moutanabbir}},
  \bibinfo{author}{\bibfnamefont{D.}~\bibnamefont{Isheim}},
  \bibinfo{author}{\bibfnamefont{D.~N.} \bibnamefont{Seidman}},
  \bibinfo{author}{\bibfnamefont{Y.}~\bibnamefont{Kawamura}}, \bibnamefont{and}
  \bibinfo{author}{\bibfnamefont{K.~M.} \bibnamefont{Itoh}},
  \bibinfo{journal}{Appl. Phys. Lett.} \textbf{\bibinfo{volume}{98}},
  \bibinfo{pages}{013111} (\bibinfo{year}{2011}).

\bibitem[{\citenamefont{Lai and Yang}(2015)}]{2015_PRB_Lai}
\bibinfo{author}{\bibfnamefont{W.~X.} \bibnamefont{Lai}} \bibnamefont{and}
  \bibinfo{author}{\bibfnamefont{W.}~\bibnamefont{Yang}},
  \bibinfo{journal}{Phys. Rev. B} \textbf{\bibinfo{volume}{92}},
  \bibinfo{pages}{155433} (\bibinfo{year}{2015}).

\bibitem[{\citenamefont{Philippopoulos
  et~al.}(2020)\citenamefont{Philippopoulos, Chesi, and Coish}}]{Perry_arXiv}
\bibinfo{author}{\bibfnamefont{P.}~\bibnamefont{Philippopoulos}},
  \bibinfo{author}{\bibfnamefont{S.}~\bibnamefont{Chesi}}, \bibnamefont{and}
  \bibinfo{author}{\bibfnamefont{W.~A.} \bibnamefont{Coish}},
  \bibinfo{journal}{arXiv:2001.05963}  (\bibinfo{year}{2020}).

\bibitem[{\citenamefont{Shekhter et~al.}(2003)\citenamefont{Shekhter, Galperin,
  Gorelik, Isacsson, and Jonson}}]{2003_J_Phys_Cond_Mat_Shekhter}
\bibinfo{author}{\bibfnamefont{R.~I.} \bibnamefont{Shekhter}},
  \bibinfo{author}{\bibfnamefont{Y.}~\bibnamefont{Galperin}},
  \bibinfo{author}{\bibfnamefont{L.~Y.} \bibnamefont{Gorelik}},
  \bibinfo{author}{\bibfnamefont{A.}~\bibnamefont{Isacsson}}, \bibnamefont{and}
  \bibinfo{author}{\bibfnamefont{M.}~\bibnamefont{Jonson}},
  \bibinfo{journal}{J. Phys.: Condens. Matter} \textbf{\bibinfo{volume}{15}},
  \bibinfo{pages}{R441} (\bibinfo{year}{2003}).

\bibitem[{\citenamefont{Angerer et~al.}(2018)\citenamefont{Angerer, Streltsov,
  Astner, Putz, Sumiya, Onoda, Isoya, Munro, Nemoto, Schmiedmayer
  et~al.}}]{2018_Nat_Phys_Angerer}
\bibinfo{author}{\bibfnamefont{A.}~\bibnamefont{Angerer}},
  \bibinfo{author}{\bibfnamefont{K.}~\bibnamefont{Streltsov}},
  \bibinfo{author}{\bibfnamefont{T.}~\bibnamefont{Astner}},
  \bibinfo{author}{\bibfnamefont{S.}~\bibnamefont{Putz}},
  \bibinfo{author}{\bibfnamefont{H.}~\bibnamefont{Sumiya}},
  \bibinfo{author}{\bibfnamefont{S.}~\bibnamefont{Onoda}},
  \bibinfo{author}{\bibfnamefont{J.}~\bibnamefont{Isoya}},
  \bibinfo{author}{\bibfnamefont{W.~J.} \bibnamefont{Munro}},
  \bibinfo{author}{\bibfnamefont{K.}~\bibnamefont{Nemoto}},
  \bibinfo{author}{\bibfnamefont{J.}~\bibnamefont{Schmiedmayer}},
  \bibnamefont{et~al.}, \bibinfo{journal}{Nat. Phys.}
  \textbf{\bibinfo{volume}{14}}, \bibinfo{pages}{1168} (\bibinfo{year}{2018}).

\bibitem[{\citenamefont{Molmer et~al.}(1993)\citenamefont{Molmer, Castin, and
  Dalibard}}]{1993_JOSAB_Molmer}
\bibinfo{author}{\bibfnamefont{K.}~\bibnamefont{Molmer}},
  \bibinfo{author}{\bibfnamefont{Y.}~\bibnamefont{Castin}}, \bibnamefont{and}
  \bibinfo{author}{\bibfnamefont{J.}~\bibnamefont{Dalibard}},
  \bibinfo{journal}{J. Opt. Soc. Am. B} \textbf{\bibinfo{volume}{10}},
  \bibinfo{pages}{524} (\bibinfo{year}{1993}).

\bibitem[{\citenamefont{Yamamoto and Imamoglu}(1999)}]{1999_BOOK_Yamamoto}
\bibinfo{author}{\bibfnamefont{Y.}~\bibnamefont{Yamamoto}} \bibnamefont{and}
  \bibinfo{author}{\bibfnamefont{A.}~\bibnamefont{Imamoglu}},
  \emph{\bibinfo{title}{Mesoscopic Quantum Optics}} (\bibinfo{publisher}{John
  Wiley and Sons, New York}, \bibinfo{year}{1999}).

\bibitem[{\citenamefont{Milton and Stegun}(1970)}]{1970_BOOK_Abramowitz}
\bibinfo{author}{\bibfnamefont{A.}~\bibnamefont{Milton}} \bibnamefont{and}
  \bibinfo{author}{\bibfnamefont{I.~A.} \bibnamefont{Stegun}},
  \emph{\bibinfo{title}{Handbook of Mathematical Functions: with Formulas,
  Graphs, and Mathematical Tables}} (\bibinfo{publisher}{Dover Publications,
  New York}, \bibinfo{year}{1970}).

\bibitem[{\citenamefont{Tokura et~al.}(2006)\citenamefont{Tokura, van~der Wiel,
  Obata, and Tarucha}}]{2006_PRL_Tokura}
\bibinfo{author}{\bibfnamefont{Y.}~\bibnamefont{Tokura}},
  \bibinfo{author}{\bibfnamefont{W.~G.} \bibnamefont{van~der Wiel}},
  \bibinfo{author}{\bibfnamefont{T.}~\bibnamefont{Obata}}, \bibnamefont{and}
  \bibinfo{author}{\bibfnamefont{S.}~\bibnamefont{Tarucha}},
  \bibinfo{journal}{Phys. Rev. Lett.} \textbf{\bibinfo{volume}{96}},
  \bibinfo{pages}{047202} (\bibinfo{year}{2006}).

\bibitem[{\citenamefont{Pioro-Ladriere
  et~al.}(2008)\citenamefont{Pioro-Ladriere, Obata, Tokura, Shin, Kubo,
  Yoshida, T., and Tarucha}}]{2008_Nat_Phys_Pioro}
\bibinfo{author}{\bibfnamefont{M.}~\bibnamefont{Pioro-Ladriere}},
  \bibinfo{author}{\bibfnamefont{T.}~\bibnamefont{Obata}},
  \bibinfo{author}{\bibnamefont{Tokura}}, \bibinfo{author}{\bibfnamefont{Y.-S.}
  \bibnamefont{Shin}}, \bibinfo{author}{\bibfnamefont{T.}~\bibnamefont{Kubo}},
  \bibinfo{author}{\bibfnamefont{K.}~\bibnamefont{Yoshida}},
  \bibinfo{author}{\bibfnamefont{T.}~\bibnamefont{T.}}, \bibnamefont{and}
  \bibinfo{author}{\bibfnamefont{S.}~\bibnamefont{Tarucha}},
  \bibinfo{journal}{Nat. Phys.} \textbf{\bibinfo{volume}{4}},
  \bibinfo{pages}{776} (\bibinfo{year}{2008}).

\bibitem[{\citenamefont{Chesi et~al.}(2014)\citenamefont{Chesi, Wang, Yoneda,
  Otsuka, Tarucha, and Loss}}]{2014_PRB_Chesi}
\bibinfo{author}{\bibfnamefont{S.}~\bibnamefont{Chesi}},
  \bibinfo{author}{\bibfnamefont{Y.-D.} \bibnamefont{Wang}},
  \bibinfo{author}{\bibfnamefont{J.}~\bibnamefont{Yoneda}},
  \bibinfo{author}{\bibfnamefont{T.}~\bibnamefont{Otsuka}},
  \bibinfo{author}{\bibfnamefont{S.}~\bibnamefont{Tarucha}}, \bibnamefont{and}
  \bibinfo{author}{\bibfnamefont{D.}~\bibnamefont{Loss}},
  \bibinfo{journal}{Phys. Rev. B} \textbf{\bibinfo{volume}{90}},
  \bibinfo{pages}{235311} (\bibinfo{year}{2014}).

\bibitem[{\citenamefont{Schmidt et~al.}(2000)\citenamefont{Schmidt, Ferrand,
  Molenkamp, Filip, and van Wees B.~J.}}]{2000_PRB_Schmidt}
\bibinfo{author}{\bibfnamefont{G.}~\bibnamefont{Schmidt}},
  \bibinfo{author}{\bibfnamefont{D.}~\bibnamefont{Ferrand}},
  \bibinfo{author}{\bibfnamefont{L.~W.} \bibnamefont{Molenkamp}},
  \bibinfo{author}{\bibfnamefont{A.~T.} \bibnamefont{Filip}}, \bibnamefont{and}
  \bibinfo{author}{\bibnamefont{van Wees B.~J.}}, \bibinfo{journal}{Phys. Rev.
  B} \textbf{\bibinfo{volume}{62}}, \bibinfo{pages}{R4790}
  (\bibinfo{year}{2000}).

\bibitem[{\citenamefont{Jasen}(2003)}]{2003_JPD_Jasen}
\bibinfo{author}{\bibfnamefont{R.}~\bibnamefont{Jasen}}, \bibinfo{journal}{J.
  Phys. D: Appl. Phys.} \textbf{\bibinfo{volume}{36}}, \bibinfo{pages}{R289}
  (\bibinfo{year}{2003}).

\bibitem[{\citenamefont{Zutic et~al.}(2006)\citenamefont{Zutic, Fabian, and
  Erwin}}]{2006_PRL_Zutic}
\bibinfo{author}{\bibfnamefont{I.}~\bibnamefont{Zutic}},
  \bibinfo{author}{\bibfnamefont{J.}~\bibnamefont{Fabian}}, \bibnamefont{and}
  \bibinfo{author}{\bibfnamefont{S.~C.} \bibnamefont{Erwin}},
  \bibinfo{journal}{Phys. Rev. Lett.} \textbf{\bibinfo{volume}{97}},
  \bibinfo{pages}{026602} (\bibinfo{year}{2006}).

\bibitem[{\citenamefont{Aurich et~al.}(2010)\citenamefont{Aurich, Baumgartner,
  Freitag, Eichler, Trbovic, and Schonenberger}}]{2010_APL_Aurich}
\bibinfo{author}{\bibfnamefont{H.}~\bibnamefont{Aurich}},
  \bibinfo{author}{\bibfnamefont{A.}~\bibnamefont{Baumgartner}},
  \bibinfo{author}{\bibfnamefont{F.}~\bibnamefont{Freitag}},
  \bibinfo{author}{\bibfnamefont{A.}~\bibnamefont{Eichler}},
  \bibinfo{author}{\bibfnamefont{J.}~\bibnamefont{Trbovic}}, \bibnamefont{and}
  \bibinfo{author}{\bibfnamefont{C.}~\bibnamefont{Schonenberger}},
  \bibinfo{journal}{Appl. Phys. Lett.} \textbf{\bibinfo{volume}{97}},
  \bibinfo{pages}{153116} (\bibinfo{year}{2010}).

\bibitem[{\citenamefont{Tarun et~al.}(2011)\citenamefont{Tarun, Huang, Fukuma,
  Idzuchi, Otani, Fukata, Ishibashi, and Oda}}]{2011_JAP_Tarun}
\bibinfo{author}{\bibfnamefont{J.}~\bibnamefont{Tarun}},
  \bibinfo{author}{\bibfnamefont{S.-Y.} \bibnamefont{Huang}},
  \bibinfo{author}{\bibfnamefont{Y.}~\bibnamefont{Fukuma}},
  \bibinfo{author}{\bibfnamefont{H.}~\bibnamefont{Idzuchi}},
  \bibinfo{author}{\bibfnamefont{Y.-C.} \bibnamefont{Otani}},
  \bibinfo{author}{\bibfnamefont{N.}~\bibnamefont{Fukata}},
  \bibinfo{author}{\bibfnamefont{K.}~\bibnamefont{Ishibashi}},
  \bibnamefont{and} \bibinfo{author}{\bibfnamefont{S.}~\bibnamefont{Oda}},
  \bibinfo{journal}{J. Appl. Phys.} \textbf{\bibinfo{volume}{109}},
  \bibinfo{pages}{07C508} (\bibinfo{year}{2011}).

\bibitem[{\citenamefont{de~Sousa and Das~Sarma}(2003)}]{2003_PRB_Sousa}
\bibinfo{author}{\bibfnamefont{R.}~\bibnamefont{de~Sousa}} \bibnamefont{and}
  \bibinfo{author}{\bibfnamefont{S.}~\bibnamefont{Das~Sarma}},
  \bibinfo{journal}{Phys. Rev. B} \textbf{\bibinfo{volume}{68}},
  \bibinfo{pages}{115322} (\bibinfo{year}{2003}).

\bibitem[{\citenamefont{George et~al.}(2010)\citenamefont{George, Witzel,
  Riemann, Abrosimov, N\"otzel, Thewalt, and Morton}}]{2010_PRL_George}
\bibinfo{author}{\bibfnamefont{R.~E.} \bibnamefont{George}},
  \bibinfo{author}{\bibfnamefont{W.}~\bibnamefont{Witzel}},
  \bibinfo{author}{\bibfnamefont{H.}~\bibnamefont{Riemann}},
  \bibinfo{author}{\bibfnamefont{N.~V.} \bibnamefont{Abrosimov}},
  \bibinfo{author}{\bibfnamefont{N.}~\bibnamefont{N\"otzel}},
  \bibinfo{author}{\bibfnamefont{M.~L.~W.} \bibnamefont{Thewalt}},
  \bibnamefont{and} \bibinfo{author}{\bibfnamefont{J.~J.~L.}
  \bibnamefont{Morton}}, \bibinfo{journal}{Phys. Rev. Lett.}
  \textbf{\bibinfo{volume}{105}}, \bibinfo{pages}{067601}
  (\bibinfo{year}{2010}).

\bibitem[{\citenamefont{Morley et~al.}(2010)\citenamefont{Morley, Warner,
  Stoneham, Greenland, van Tol, Kay, and Aeppli}}]{2010_Nat_Mater_Morley}
\bibinfo{author}{\bibfnamefont{G.~W.} \bibnamefont{Morley}},
  \bibinfo{author}{\bibfnamefont{M.}~\bibnamefont{Warner}},
  \bibinfo{author}{\bibfnamefont{A.~M.} \bibnamefont{Stoneham}},
  \bibinfo{author}{\bibfnamefont{P.~T.} \bibnamefont{Greenland}},
  \bibinfo{author}{\bibfnamefont{J.}~\bibnamefont{van Tol}},
  \bibinfo{author}{\bibfnamefont{C.~W.~M.} \bibnamefont{Kay}},
  \bibnamefont{and} \bibinfo{author}{\bibfnamefont{G.}~\bibnamefont{Aeppli}},
  \bibinfo{journal}{Nat. Mater.} \textbf{\bibinfo{volume}{9}},
  \bibinfo{pages}{725} (\bibinfo{year}{2010}).

\bibitem[{\citenamefont{Mourik et~al.}(2018)\citenamefont{Mourik, Asaad,
  Firgau, Pla, Holmes, Milburn, McCallum, and Morello}}]{2018_PRE_Mourik}
\bibinfo{author}{\bibfnamefont{V.}~\bibnamefont{Mourik}},
  \bibinfo{author}{\bibfnamefont{S.}~\bibnamefont{Asaad}},
  \bibinfo{author}{\bibfnamefont{H.}~\bibnamefont{Firgau}},
  \bibinfo{author}{\bibfnamefont{J.~J.} \bibnamefont{Pla}},
  \bibinfo{author}{\bibfnamefont{C.}~\bibnamefont{Holmes}},
  \bibinfo{author}{\bibfnamefont{G.~J.} \bibnamefont{Milburn}},
  \bibinfo{author}{\bibfnamefont{J.~C.} \bibnamefont{McCallum}},
  \bibnamefont{and} \bibinfo{author}{\bibfnamefont{A.}~\bibnamefont{Morello}},
  \bibinfo{journal}{Phys. Rev. E} \textbf{\bibinfo{volume}{98}},
  \bibinfo{pages}{042206} (\bibinfo{year}{2018}).

\bibitem[{\citenamefont{Asaad et~al.}(2019)\citenamefont{Asaad, Mourik,
  Joecker, Johnson, Baczewski, Firgau, Madzik, Schmitt, Pla, Hudson
  et~al.}}]{2019_arXiv_Asaad}
\bibinfo{author}{\bibfnamefont{S.}~\bibnamefont{Asaad}},
  \bibinfo{author}{\bibfnamefont{V.}~\bibnamefont{Mourik}},
  \bibinfo{author}{\bibfnamefont{B.}~\bibnamefont{Joecker}},
  \bibinfo{author}{\bibfnamefont{M.~A.~I.} \bibnamefont{Johnson}},
  \bibinfo{author}{\bibfnamefont{A.~D.} \bibnamefont{Baczewski}},
  \bibinfo{author}{\bibfnamefont{H.~R.} \bibnamefont{Firgau}},
  \bibinfo{author}{\bibfnamefont{M.~T.} \bibnamefont{Madzik}},
  \bibinfo{author}{\bibfnamefont{V.}~\bibnamefont{Schmitt}},
  \bibinfo{author}{\bibfnamefont{J.~J.} \bibnamefont{Pla}},
  \bibinfo{author}{\bibfnamefont{F.~E.} \bibnamefont{Hudson}},
  \bibnamefont{et~al.}, \bibinfo{journal}{arXiv:1906.01086v1}
  (\bibinfo{year}{2019}).

\bibitem[{\citenamefont{Franke et~al.}(2015)\citenamefont{Franke, Hrubesch,
  K\"unzl, Becker, Itoh, Stutzmann, Hoehne, Dreher, and
  Brandt}}]{2015_PRL_Franke}
\bibinfo{author}{\bibfnamefont{D.~P.} \bibnamefont{Franke}},
  \bibinfo{author}{\bibfnamefont{F.~M.} \bibnamefont{Hrubesch}},
  \bibinfo{author}{\bibfnamefont{M.}~\bibnamefont{K\"unzl}},
  \bibinfo{author}{\bibfnamefont{H.-W.} \bibnamefont{Becker}},
  \bibinfo{author}{\bibfnamefont{K.~M.} \bibnamefont{Itoh}},
  \bibinfo{author}{\bibfnamefont{M.}~\bibnamefont{Stutzmann}},
  \bibinfo{author}{\bibfnamefont{F.}~\bibnamefont{Hoehne}},
  \bibinfo{author}{\bibfnamefont{L.}~\bibnamefont{Dreher}}, \bibnamefont{and}
  \bibinfo{author}{\bibfnamefont{M.~S.} \bibnamefont{Brandt}},
  \bibinfo{journal}{Phys. Rev. Lett.} \textbf{\bibinfo{volume}{115}},
  \bibinfo{pages}{057601} (\bibinfo{year}{2015}).

\bibitem[{\citenamefont{Pla et~al.}(2018)\citenamefont{Pla, Bienfait, Pica,
  Mansir, Mohiyaddin, Zeng, Niquet, Morello, Schenkel, Morton
  et~al.}}]{2018_PRAppl_Pla}
\bibinfo{author}{\bibfnamefont{J.~J.} \bibnamefont{Pla}},
  \bibinfo{author}{\bibfnamefont{A.}~\bibnamefont{Bienfait}},
  \bibinfo{author}{\bibfnamefont{G.}~\bibnamefont{Pica}},
  \bibinfo{author}{\bibfnamefont{J.}~\bibnamefont{Mansir}},
  \bibinfo{author}{\bibfnamefont{F.~A.} \bibnamefont{Mohiyaddin}},
  \bibinfo{author}{\bibfnamefont{Z.}~\bibnamefont{Zeng}},
  \bibinfo{author}{\bibfnamefont{Y.~M.} \bibnamefont{Niquet}},
  \bibinfo{author}{\bibfnamefont{A.}~\bibnamefont{Morello}},
  \bibinfo{author}{\bibfnamefont{T.}~\bibnamefont{Schenkel}},
  \bibinfo{author}{\bibfnamefont{J.~J.~L.} \bibnamefont{Morton}},
  \bibnamefont{et~al.}, \bibinfo{journal}{Phys. Rev. Applied}
  \textbf{\bibinfo{volume}{9}}, \bibinfo{pages}{044014} (\bibinfo{year}{2018}).

\bibitem[{\citenamefont{Mansir et~al.}(2018)\citenamefont{Mansir, Conti, Zeng,
  Pla, Bertet, Swift, Van~de Walle, Thewalt, Sklenard, Niquet
  et~al.}}]{2018_PRL_Mansir}
\bibinfo{author}{\bibfnamefont{J.}~\bibnamefont{Mansir}},
  \bibinfo{author}{\bibfnamefont{P.}~\bibnamefont{Conti}},
  \bibinfo{author}{\bibfnamefont{Z.}~\bibnamefont{Zeng}},
  \bibinfo{author}{\bibfnamefont{J.~J.} \bibnamefont{Pla}},
  \bibinfo{author}{\bibfnamefont{P.}~\bibnamefont{Bertet}},
  \bibinfo{author}{\bibfnamefont{M.~W.} \bibnamefont{Swift}},
  \bibinfo{author}{\bibfnamefont{C.~G.} \bibnamefont{Van~de Walle}},
  \bibinfo{author}{\bibfnamefont{M.~L.~W.} \bibnamefont{Thewalt}},
  \bibinfo{author}{\bibfnamefont{B.}~\bibnamefont{Sklenard}},
  \bibinfo{author}{\bibfnamefont{Y.~M.} \bibnamefont{Niquet}},
  \bibnamefont{et~al.}, \bibinfo{journal}{Phys. Rev. Lett.}
  \textbf{\bibinfo{volume}{120}}, \bibinfo{pages}{167701}
  (\bibinfo{year}{2018}).

\bibitem[{\citenamefont{Breuer and Petruccione}(2002)}]{2002_BOOK_Breuer}
\bibinfo{author}{\bibfnamefont{H.-P.} \bibnamefont{Breuer}} \bibnamefont{and}
  \bibinfo{author}{\bibfnamefont{F.}~\bibnamefont{Petruccione}},
  \emph{\bibinfo{title}{The Theory of Open Quantum Systems}}
  (\bibinfo{publisher}{Oxford University Press, New York},
  \bibinfo{year}{2002}).

\bibitem[{\citenamefont{Blum}(2012)}]{2012_BOOK_Blum}
\bibinfo{author}{\bibfnamefont{K.}~\bibnamefont{Blum}},
  \emph{\bibinfo{title}{Density Matrix Theory and Applications}}
  (\bibinfo{publisher}{Springer, Heidelberg}, \bibinfo{year}{2012}).

\end{thebibliography}

\end{document}